\documentclass[sigconf,natbib=true,nonacm]{acmart}
\usepackage{graphicx}
\usepackage{multicol}
\usepackage{multirow}
\usepackage{amsfonts}
\usepackage{bm}
\usepackage{amsmath}
\usepackage{amsthm}
\usepackage{hhline}

\usepackage{colortbl}
\usepackage{xcolor}
\usepackage{booktabs}
\usepackage{arydshln}  %
\usepackage{float}

\definecolor{bluegrey}{RGB}{230, 240, 255}

\usepackage{pifont}

\usepackage{listings}
\lstdefinestyle{prompt}{
  basicstyle=\ttfamily\scriptsize,
  breaklines=true,
  breakatwhitespace=false,
  columns=fullflexible,
  keepspaces=true,
  showstringspaces=false,
  frame=single,
  framesep=4pt,
  xleftmargin=2pt,
  aboveskip=4pt,
  belowskip=4pt,
}

\definecolor{citecolor}{RGB}{34,139,34} 
\hypersetup{breaklinks=true,colorlinks=true,
citecolor=citecolor,bookmarks=false}

\definecolor{bittersweet}{rgb}{1.0, 0.44, 0.37}

\definecolor{mygreen}{rgb}{0.29, 0.7, 0.48}

\definecolor{mygray}{gray}{0.4}

\AtBeginDocument{%
  \providecommand\BibTeX{{%
    \normalfont B\kern-0.5em{\scshape i\kern-0.25em b}\kern-0.8em\TeX}}}

\setcopyright{none}
\begin{document}

\title{RALI@iKAT 2026: Retrieval-aware User Profile Denoising via Personalized Conversational Dense Retrieval}

\author{Yuchen Hui}
\orcid{0000-0002-9659-3714}
\affiliation{%
  \institution{RALI, Université de Montréal}
  \city{Montréal}
  \state{Québec}
  \country{Canada}
}
\email{yuchen.hui@umontreal.ca}

\author{Fengran Mo}
\orcid{0000-0002-0838-6994}
\affiliation{%
  \institution{RALI, Université de Montréal}
  \city{Montréal}
  \state{Québec}
  \country{Canada}
}
\email{fengran.mo@umontreal.ca}

\author{Jia Ao Sun}
\affiliation{%
  \institution{RALI, Université de Montréal}
  \city{Montréal}
  \state{Québec}
  \country{Canada}
}
\email{jia.ao.sun@umontreal.ca}

\author{Yuxing Tian}
\affiliation{%
  \institution{RALI, Université de Montréal}
  \city{Montréal}
  \state{Québec}
  \country{Canada}
}
\email{yuxing.tian@umontreal.ca}

\author{Wenhao Xu}
\orcid{0009-0002-8110-0849}
\affiliation{%
  \institution{RALI, Université de Montréal}
  \city{Montréal}
  \state{Québec}
  \country{Canada}
}
\email{wenhao.xu@umontreal.ca}

\author{Jian-Yun Nie}
\orcid{0000-0003-1556-3335}
\affiliation{%
  \institution{RALI, Université de Montréal}
  \city{Montréal}
  \state{Québec}
  \country{Canada}
}
\email{nie@iro.umontreal.ca}

\renewcommand{\shortauthors}{Hui et al.}

\begin{abstract}
Modern conversational information access systems continuously extract user information from conversations to build user profiles for personalized retrieval. As user profiles accumulate over time, they inevitably become noisy, making effective user profile denoising a fundamental challenge. Existing personalized conversational query reformulation methods rely on the black-box reasoning capability of LLMs to expand the original query with relevant user profile phrases, producing a verbose query for retrieval. Such rewrite-then-retrieve pipelines reduce user profile denoising to the explicit selection of profile phrases in the sparse token space, separating personalization from downstream relevance modeling. Consequently, they often fail to identify user information that is truly beneficial for retrieval.

To enable retrieval-aware user profile denoising, we propose Personalized Conversational Dense Retrieval (PCDR), which learns to jointly project the conversation history and user profile into a latent embedding space dedicated to retrieval. By inheriting the context denoising capability from conversational dense retrievers, PCDR requires only a handful of personalized relevance annotations for training. We evaluate PCDR on the iKAT 2026 tasks, where it serves as the core retrieval component of a conversational AI agent assisting users with complex information needs.

\end{abstract}




\maketitle

\section{Introduction}
\label{sec:intro}
\begin{figure}[t]
  \centering
  \includegraphics[width=\columnwidth]{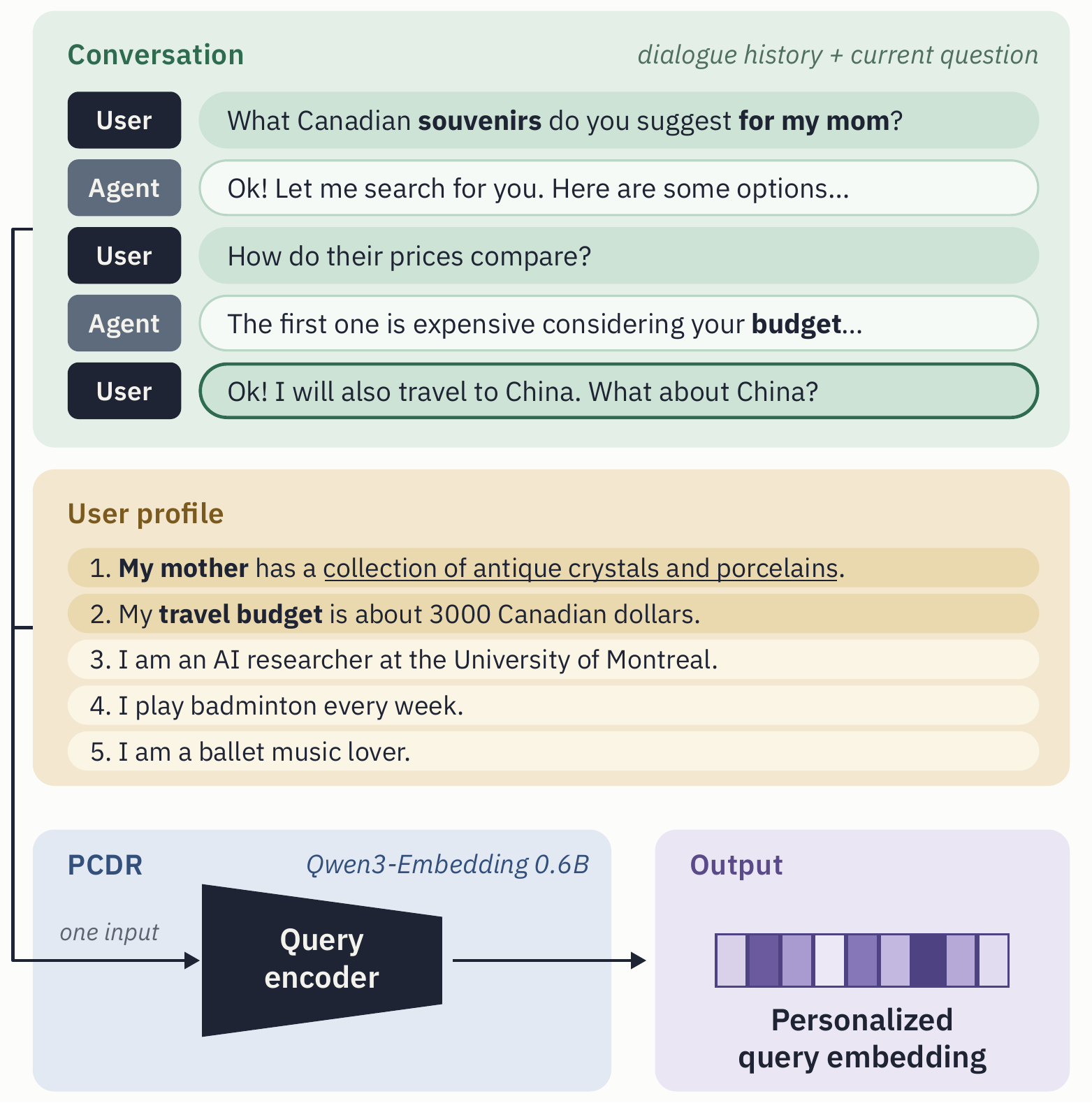}
  \caption{An example of personalized conversational search. The current question depends on the dialogue history and on only two of the five profile statements, which are highlighted. PCDR encodes the conversation and the full profile into a single personalized query embedding.}
  \label{fig:example}
\end{figure}

Conversational Search systems resolve a user's information need through multiple turns of natural-language interaction~\cite{gao2022neural,Zamani2022ConversationalIS}. Personalized Conversational Search systems additionally condition conversational retrieval on a profile of persistent facts about the user. The TREC Interactive Knowledge Assistance Track (iKAT)~\cite{aliannejadi2024ikat} targets this problem by providing human-curated information-seeking conversations accompanied by user profiles consisting of natural-language statements. Figure~\ref{fig:example} shows an example: the current question depends on the dialogue history and on only a few statements of the user profile, and PCDR encodes the conversation and the full profile into a single personalized query embedding.

This year, iKAT 2026 introduced an interactive task that requires the participating retrieval system to interact with the same simulated user~\cite{owoicho2023exploiting} across multiple conversation sessions. Submitted systems should continuously extract new persistent user facts and append them to the profile throughout these sequential interactions. The profile therefore grows monotonically as the dialogue unfolds.

This accumulation comes at a cost. Long-term memory mechanisms for conversational agents are typically designed to add new information more readily than they consolidate or remove it~\cite{Zhang2024,Park2023}. Over extended interactions, the profile becomes long and accumulates statements that are only loosely related to any individual question. The resulting profile is informative but noisy: information that is decisive for one question may distract retrieval for the next, and the relevant profile statements cannot be determined in advance.

This problem is amplified by the prevailing approach for exploiting user profiles: Personalized Conversational Query Reformulation (PCQR)~\cite{mo2024leverage,mo2025apcir,lupart2024irlab,hui2024rali,dong2025personalise,dong2026apr}.
Given the full conversation history and the user profile, PCQR employs Large Language Models (LLMs) to generate a personalized and de-contextualized self-contained query to submit to off-the-shelf ad-hoc retrievers. However, previous studies~\cite{mo2025apcir, hui2024rali} showed that this process can cause \emph{over-personalization} query drift, where LLMs incorporate weakly related profile phrases into the rewritten query. As profiles grow through continued interaction, this problem becomes increasingly severe.

We argue that this limitation arises from the intermediate rewritten query. First, personalization must be verbalized through explicit profile phrase injection in the discrete token space, which cannot fully express implicit, retrieval-relevant meanings. For example, a user's interest in \emph{technology} may imply \emph{science fiction} when searching for books or \emph{futuristic} when searching for clothing. Second, queries generated by zero-shot LLMs are not directly optimized for retrieval. Even when the selected profile phrases are relevant, inserting them into the query does not necessarily produce a good search query.

\begin{figure}[t]
  \centering
  \includegraphics[width=\columnwidth]{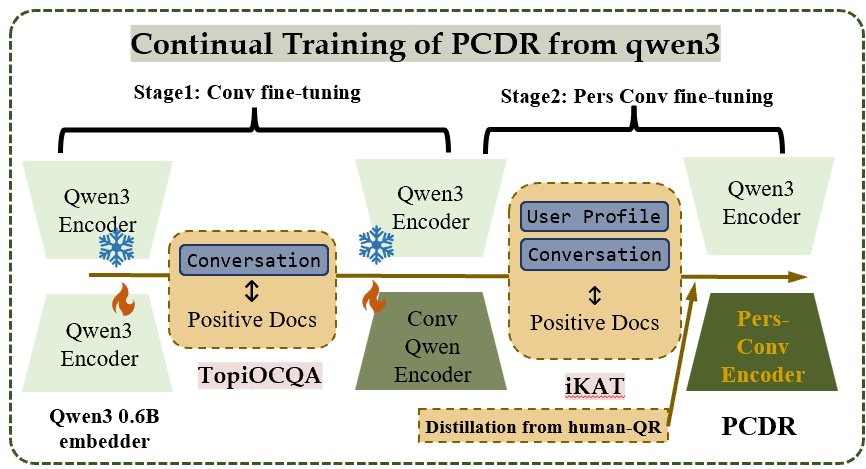}
  \caption{Two-stage continual training of PCDR. The query encoder of Qwen3-Embedding 0.6B is first fine-tuned on TopiOCQA conversations to obtain conv-Qwen3, and then on iKAT conversations and user profiles, with distillation from human query rewrites, to obtain PCDR. The document encoder stays frozen in both stages.}
  \label{fig:continual_pcdr}
\end{figure}

To overcome these limitations, we propose \textbf{Personalized Conversational Dense Retrieval (PCDR)}, which performs retrieval-aware user profile denoising directly in the dense retriever's latent embedding space. PCDR jointly encodes the conversation history and user profile into a single query embedding, allowing implicit personal information to influence retrieval without being explicitly verbalized in a rewritten query. Trained with a small number of personalized relevance judgments from the iKAT 2023--2025 datasets, PCDR learns to condition the encoding of the contextual dense embedding only on the relevant parts of the user profile that are also useful for search. This can be viewed as an implicit user profile denoising process that is optimized directly for the retrieval task.  We further distill knowledge from personalized human rewrites to stabilize training. Figure~\ref{fig:continual_pcdr} illustrates the two-stage continual training that produces PCDR from a base Qwen3 0.6B embedding model. Visualization analysis illustrates that PCDR focuses on fewer but more essential profile signals than explicit profile phrase selection.

We evaluate PCDR by incorporating it in a search system participating in the online interactive track of iKAT 2026. The results show that for complex real-world information needs, PCDR alone produces effective personalized retrieval results, while combining it with heterogeneous retrievers provides further improvements.

\section{Related Work}
\label{sec:related}

\noindent\textbf{Conversational Dense Retrieval.}
Conversational search aims at resolving context-dependent utterances against a long and noisy dialogue history~\cite{mo2025survey}. Instead of rewriting the utterance into a self-contained query~\cite{mo2023convgqr,mao2023large,mo2024chiq}, conversational dense retrieval (CDR) encodes the whole session into a single query embedding, learning to denoise the conversation in latent space and to keep only the context that reveals the current search intent~\cite{qu2020open,yu2021few,lin2021contextualized}.
CDR has since been refined in many directions~\cite{yu2021few,kim2022saving,mao2022curriculum,mao2022convtrans,mao2023learning,mo2023learning,jin2023instructor,mo2024history,mo2024aligning,chen2024generalizing,mo2024convsdg,mo2026convmix,lupart2025disco} and extended to LLM-based embedders~\cite{mao2024chatretriever,mo2025uniconv,yang2025contextual,son2026rcem,jiang2026onemodel}. PCDR is continually fine-tuned from a CDR checkpoint, so that it inherits the ability to denoise a long conversational context and can apply it to the user profile, preventing the profile from overloading the search intent with irrelevant information.

\noindent\textbf{Personalized Dense Retrieval.}
A similar idea is widely used in personalized dense retrieval, where various forms of user information are jointly encoded with the query: latent or ID-based user embeddings~\cite{ai2017hem,ai2019zam,bi2020tem,bao2024pr2}, categorical profile features~\cite{huang2020ebr,zhang2020dpsr}, behavioral histories such as past queries, clicks, and purchases~\cite{li2021mgdspr,jha2024etsy,belyi2023personalized,zeng2023uia,jiang2026ttp}, and structured textual constraints~\cite{wang2026skillfeed}. The retrieval targets range from products and videos to agent skills. Instruction-following retrievers have also been studied with long-form personalized instructions~\cite{weller2025promptriever,oh2024instructir}. In these tasks, although complicated, the information need is still clearly stated by the instructions, not demanding as much reasoning effort as inferring the search intent implied by a stand-alone user profile. 

None of these works conditions large-scale web passage retrieval on a persistent, unstructured natural-language profile of a specific user together with that user's conversational history. Such natural human--agent interactions are noisier, and their personal intents are often only implied by the interactions of phrasing nuances in both profiles and dialogues, which makes dense retrieval considerably harder.

\noindent\textbf{Personalized Conversational Information Retrieval.}
Personalized conversational information retrieval (PCIR) studies exactly this setting~\cite{aliannejadi2024ikat,abbasiantaeb2026ikat}. Existing systems mostly follow the Personalized Conversational Query Reformulation (PCQR) paradigm: they rely on the implicit incorporation of profile statements based on the parametric knowledge of LLMs to produce a plausibly personalized search query, then pass it to off-the-shelf retrievers and rerankers~\cite{mo2024leverage,mo2025apcir,hui2024rali,lupart2024irlab,dong2025personalise,dong2026apr}. Since such LLM-based user profile denoising that is not trained in favor of retrieval can be imperfect~\cite{dong2026apr}, PCDR instead trains a retrieval model to denoise the user profile directly.

\section{Task Formulation}
\label{sec:task}
We address personalized conversational search over a fixed passage corpus $\mathcal{C}$. A conversation proceeds over turns $t=1,2,\dots$; at turn $t$ the user issues an utterance $u_t$ whose information need may depend on the dialogue history $H_t=(u_1,r_1,\dots,u_{t-1},r_{t-1})$, where $r_i$ is the system response at turn $i$. The user is characterized by a profile $P_t=\{p_1,\dots,p_{m_t}\}$, a set of natural-language statements about the user, for instance ``\textit{I am vegetarian}''. Appendix~\ref{app:sample} shows an example of a conversation and its user profile. Given $(u_t, H_t, P_t)$, the system returns a ranking of passages from $\mathcal{C}$ that satisfy the information need behind $u_t$ and are consistent with the profile $P_t$, and then generates a response $r_t$ grounded in the top-ranked passages.

The profile is dynamic in the interactive setting: it is not supplied in advance but accumulated online. After turn $t$, new durable facts inferred from $(u_t,r_t)$ are appended, so $P_{t}\subseteq P_{t+1}$ and $|P_t|$ grows monotonically across turns and across a user's successive sessions. In the offline training and validation setting of Sections~\ref{sec:method} and~\ref{sec:experiments}, the profile is instead the fixed, human-annotated user profile of the conversation.

\section{Method}
\label{sec:method}

\subsection{Overview}
\label{sec:overview}
Following existing practice in CDR~\cite{yu2021few,lin2021contextualized,mao2022curriculum,mo2024history,mo2024aligning}, only the query side of the PCDR dual encoder is fine-tuned, while the document encoder is kept frozen. More specifically, a query encoder $f_\theta$ encodes the current utterance together with the dialogue history and the user profile into a single vector $q=f_\theta(u_t,H_t,P_t)$, and passages are ranked by the cosine similarity $\cos(q,d)$ between $q$ and their frozen embeddings; only $\theta$ is trained. Unlike the reformulate-then-retrieve pipelines of our previous systems~\cite{hui2024rali,mo2025apcir}, PCDR forms no intermediate rewritten query and consumes the conversation and the user profile directly. We initialize $f_\theta$ from a conversational backbone, conv-Qwen3, which is described in Section~\ref{sec:backbone}. Section~\ref{sec:query} describes how the query is constructed from the conversation and the user profile, and Section~\ref{sec:loss} presents the composite training objective, which distills a personalized human rewrite into the conversational query.
The encoder produced by this recipe is PCDR. This section and Section~\ref{sec:experiments} concern training and offline evaluation, where the profile is fixed and human-annotated; Section~\ref{sec:runs} describes how PCDR is deployed in the interactive task, where the profile is accumulated turn by turn.

\subsection{Training Data}
\label{sec:data}
Having outlined the model, we now describe the data it is trained and evaluated on. PCDR is trained and validated entirely on \emph{previous-year} iKAT data and tested online on the iKAT 2026 simulated users, as reported in Section~\ref{sec:simresults}; no 2026 relevance labels are used in training.

\noindent\textbf{iKAT 2023--2025.}
We use the graded relevance judgments of the TREC Interactive Knowledge Assistance Track (iKAT)~\cite{aliannejadi2024ikat} 2023--2025, a personalized conversational benchmark over the ClueWeb22-B web corpus~\cite{overwijk2022clueweb}. Each conversation is paired with a user profile as defined in Section~\ref{sec:task}. Each turn provides the context-dependent utterance $u_t$, a human \emph{oracle rewrite} that resolves it into a self-contained query and is used only as a distillation teacher in Section~\ref{sec:loss}, and, for the relevant turns, human annotations of which profile statements are relevant. Passage relevance is graded on a $0$--$4$ scale. We use all $324$ turns that have relevance judgments: $176$ from 2023, $103$ from 2024, and $45$ from 2025.

\noindent\textbf{TopiOCQA.}
To obtain a conversational backbone we fine-tune on TopiOCQA~\cite{adlakha2022topiocqa}, an open-domain conversational passage-retrieval dataset over Wikipedia with topic shifts, which contains about $45$K history-aware query--passage pairs.

\noindent\textbf{Splits.}
We split the iKAT judged turns \emph{by conversation} with a $9{:}1$ ratio, stratified by year, into $280$ training and $44$ held-out turns, and report NDCG@$3$ on the held-out turns under full-corpus retrieval over the $116$M-passage ClueWeb22-B index. Each training turn carries on average ${\sim}177$ graded candidate passages.

\subsection{Conversational Backbone: conv-Qwen3}
\label{sec:backbone}
We first fine-tune base-Qwen3-Embedding~\cite{qwen3emb}, a $0.6$B general-purpose embedding model, on TopiOCQA to obtain a CDR model, conv-Qwen3, which serves both as a strong zero-shot baseline and as the initialization for personalized fine-tuning. The fine-tuning recipe and the criterion for selecting the backbone checkpoint are given in Section~\ref{sec:impl}.

\subsection{Query Construction}
\label{sec:query}
Following the Qwen3-Embedding protocol~\cite{qwen3emb}, a query is an instruction-prefixed string. The instruction is placed on the query side only, and documents are encoded without instruction. We consider three forms:
\begin{itemize}
  \item \textbf{full-profile}: the instruction ``\textit{Given a conversation between a user and an AI assistant and the user's profile, retrieve passages that answer the user's last question in a way consistent with the user's profile}'', followed by the \emph{entire} profile and the conversation history ending in the current question.
  \item \textbf{relevant-profile}: identical, but the profile contains only the human-annotated relevant profile statements, an oracle profile selection.
  \item \textbf{oracle rewrite}, used only as the teacher: the ad-hoc instruction ``\textit{Given a web search query, retrieve relevant passages that answer the query}'' followed by the human oracle rewrite.
\end{itemize}

\subsection{Training Objective}
\label{sec:loss}
Let $q$ be the trained query embedding and $\{d_i\}$ the candidate passages of a turn with grades $g_i\!\in\!\{0,\dots,4\}$; all embeddings are $\ell_2$-normalized.

\noindent\textbf{Graded fine-tuning.} We adopt BiXSE~\cite{tsirigotis2025bixse}, a pointwise objective that treats the normalized grade as a probabilistic target under binary cross-entropy:
\setlength{\abovedisplayskip}{3pt}\setlength{\belowdisplayskip}{3pt}%
\begin{equation}
\mathcal{L}_{\text{grade}} = \frac{1}{|\mathcal{P}|}\sum_{i\in\mathcal{P}} \mathrm{BCE}\!\big(\sigma(\alpha\cos(q,d_i)+\beta),\; g_i/4\big),
\end{equation}
with scale $\alpha{=}20$ and a learnable bias $\beta$. The pool $\mathcal{P}$ mixes the turn's own judged documents with global negatives, described below.

\noindent\textbf{Oracle-rewrite distillation.} Following the ConvDR paradigm~\cite{yu2021few} of distilling a manual rewrite into the conversational query, we pull $q$ toward the embedding $t$ of the turn's human oracle rewrite, which is encoded by the frozen base-Qwen3 and receives no gradient:
\begin{equation}
\mathcal{L}_{\text{KD}} = 1-\cos(q, t).
\end{equation}
Since $t$ carries the personalized information and lies on the frozen document manifold, KD both teaches personalization and stabilizes training.

\noindent\textbf{Initialization anchor.} To prevent $q$ from rotating off the document manifold, we anchor it to its conv-Qwen3 initialization embedding $q_0$:
\begin{equation}
\mathcal{L}_{\text{anchor}} = 1-\cos(q, q_0).
\end{equation}

\noindent\textbf{Global negatives.} Beyond a turn's judged pool, each step we sample $K$ passages from a corpus-wide negative pool and append them as grade-$0$ candidates in $\mathcal{L}_{\text{grade}}$, calibrating the encoder against the full corpus rather than only the shallow judged pool.

\noindent The total objective is
\begin{equation}
\mathcal{L} = w_{g}\,\mathcal{L}_{\text{grade}} + w_{k}\,\mathcal{L}_{\text{KD}} + w_{a}\,\mathcal{L}_{\text{anchor}}.
\end{equation}

\subsection{Implementation Details}
\label{sec:impl}
\noindent\textbf{conv-Qwen3 fine-tuning.} We fine-tune base-Qwen3-Embedding on TopiOCQA with an InfoNCE~\cite{oord2019representation} contrastive loss with a temperature of $0.01$, cross-GPU in-batch negatives, and false-negative masking, a batch size of $480$ over four GPUs, a constant learning rate $1\mathrm{e}{-5}$ with $6\%$ warmup, $20$ epochs, and fp32 master weights. \emph{Checkpoint selection:} we evaluate every epoch on the out-of-domain QReCC conversational benchmark and select the peak epoch, epoch~$1$, whose NDCG@$10$ of $54.6$ exceeds the $52.5$ of the zero-shot base model; later epochs monotonically forget, so the backbone is early-stopped at its conversational peak.

\noindent\textbf{Personalized fine-tuning.} We train the query encoder with the objective of Section~\ref{sec:loss} for up to $30$ epochs, batch size $16$, constant learning rate $5\mathrm{e}{-6}$, bf16 with fp32 master weights, keeping all judged grade-$0$ docs as negatives plus $K{=}2048$ global negatives; default weights $w_g{=}1,\,w_k{=}2,\,w_a{=}1$. Whether to use LoRA~\cite{hu2022lora} or full fine-tuning is decided empirically in Section~\ref{sec:hpsel}.

\subsection{Hyperparameter Selection}
\label{sec:hpsel}
All model selection uses \emph{full-corpus} held-out NDCG@$3$; we found a shallow pool-reranking proxy to be anti-correlated with full-corpus retrieval and therefore unsafe for selection. Table~\ref{tab:hpsel} reports the search over the adaptation method, comparing LoRA ranks with full fine-tuning, and over the learning rate, using conv-Qwen3 initialization, the full-profile query, and the complete objective. NDCG@$3$ rises with the LoRA rank and saturates at $r{=}64$, and LoRA with $r{=}64$ and a learning rate of $5\mathrm{e}{-6}$ outperforms the best full fine-tuning.

\begin{table}[t]
    \centering
    \caption{Hyperparameter selection for personalized fine-tuning, in NDCG@$3$ on the held-out turns with full-corpus retrieval. All runs start from conv-Qwen3 and use the full-profile query and the complete objective. Bold indicates the best result.}
    \vspace{-2ex}
    \begin{tabular}{llc}
    \toprule
        Adaptation & Learning rate & NDCG@3 \\
        \midrule
        \multicolumn{3}{l}{\emph{LoRA rank} (lr $5\mathrm{e}{-6}$)} \\
        \quad LoRA $r{=}4$   & $5\mathrm{e}{-6}$ & 30.6 \\
        \quad LoRA $r{=}8$   & $5\mathrm{e}{-6}$ & 31.0 \\
        \quad LoRA $r{=}16$  & $5\mathrm{e}{-6}$ & 31.2 \\
        \quad LoRA $r{=}32$  & $5\mathrm{e}{-6}$ & 32.2 \\
        \quad LoRA $r{=}64$  & $5\mathrm{e}{-6}$ & \textbf{33.1} \\
        \quad LoRA $r{=}96$  & $5\mathrm{e}{-6}$ & 32.6 \\
        \quad LoRA $r{=}128$ & $5\mathrm{e}{-6}$ & 32.8 \\
        \midrule
        \multicolumn{3}{l}{\emph{learning rate}} \\
        \quad LoRA $r{=}16$  & $1\mathrm{e}{-6}$ & 30.0 \\
        \quad LoRA $r{=}16$  & $1\mathrm{e}{-5}$ & 31.2 \\
        \quad LoRA $r{=}16$  & $2\mathrm{e}{-5}$ & 31.6 \\
        \quad LoRA $r{=}64$  & $2\mathrm{e}{-5}$ & 32.1 \\
        \midrule
        \multicolumn{3}{l}{\emph{full fine-tuning}} \\
        \quad full FT & $2\mathrm{e}{-6}$ & 30.1 \\
        \quad full FT & $5\mathrm{e}{-6}$ & 32.2 \\
        \quad full FT & $1\mathrm{e}{-5}$ & 30.0 \\
        \bottomrule
     \end{tabular}
     \label{tab:hpsel}
\end{table}

\section{Experiments}
\label{sec:experiments}

\subsection{Main Results}
Table~\ref{tab:baselines} reports the zero-shot baselines under the two query forms, with PCDR shown for reference; all numbers are NDCG@$3$ on the held-out turns with full-corpus retrieval. The first two rows apply the query encoders without any iKAT fine-tuning. The general-purpose base-Qwen3 is weak, whereas its TopiOCQA fine-tune conv-Qwen3 is far stronger without ever seeing iKAT data, confirming that conversational pre-training transfers to personalized retrieval. Table~\ref{tab:cumulative} then builds up our method on top of conv-Qwen3 with LoRA and the full-profile query, each row adding one component to the row above. Distillation alone roughly matches the conv-Qwen3 baseline; adding the graded ranking loss improves it further, the initialization anchor adds a small gain, and global negatives give the largest single jump to the full method, our PCDR model, which is our best configuration and comes closest to the oracle-rewrite upper bound. Every component contributes positively.

\begin{table}[t]
    \centering
    \caption{Zero-shot baselines and PCDR under the two query forms, in NDCG@$3$ on the held-out turns with full-corpus retrieval. Bold indicates the best result.}
    \vspace{-2ex}
    \begin{tabular}{lcc}
    \toprule
        Encoder & full-profile & relevant-profile \\
        \midrule
        base-Qwen3 & 12.9 & 13.4 \\
        conv-Qwen3 & 29.0 & 28.9 \\
        \midrule
        PCDR (ours) & \textbf{33.1} & 32.0 \\
        \bottomrule
     \end{tabular}
     \label{tab:baselines}
\end{table}

\begin{table}[t]
    \centering
    \caption{Cumulative build-up of PCDR on top of conv-Qwen3, in NDCG@$3$ on the held-out turns. Each row adds one component to the row above. The upper bound encodes the human oracle rewrite with base-Qwen3.}
    \vspace{-2ex}
    \begin{tabular}{lc}
    \toprule
        Configuration & NDCG@3 \\
        \midrule
        pure KD & 28.8 \\
        \quad + graded fine-tune (BiXSE) & 30.3 \\
        \quad + initialization anchor & 30.4 \\
        \quad + global negatives (full method, PCDR) & \textbf{33.1} \\
        \midrule
        oracle upper bound & 36.3 \\
        \bottomrule
     \end{tabular}
     \label{tab:cumulative}
\end{table}

\subsection{Ablation Study}
Table~\ref{tab:ablation} isolates each component by leave-one-out, removing one at a time from the full method while keeping everything else, including the other loss weights, fixed. The conversational backbone is the single most important ingredient, and replacing conv-Qwen3 with base-Qwen3 hurts the most by a wide margin. Oracle-rewrite distillation and the graded ranking loss come next and matter almost as much, followed by the initialization anchor and the global negatives. The remaining choices, namely the full-profile query form, LoRA over full fine-tuning, and the low learning rate, each contribute a smaller amount.

We also tried three techniques that help in other retrieval settings, namely hard-negative mining, listwise distillation, and profile-subset data augmentation, but none improves over the full method, whether applied alone or together. Hard negatives are in fact counterproductive here: mined from the oracle's own top retrievals, they are predominantly unjudged yet relevant and therefore act as false negatives. We conclude that, under a frozen document index and only a few hundred judged turns, the residual gap to the oracle upper bound reflects the intrinsic mismatch between the long, noisy full-profile conversation and the short, clean oracle rewrite.

\begin{table}[t]
    \centering
    \caption{Leave-one-out ablation of PCDR, in NDCG@$3$ on the held-out turns. Each row removes one component from the full method and keeps all other settings fixed.}
    \vspace{-2ex}
    \begin{tabular}{lc}
    \toprule
        Configuration & NDCG@3 \\
        \midrule
        full method & \textbf{33.1} \\
        \midrule
        $-$ conv-Qwen3 fine-tune (use base-Qwen3) & 27.5 \\
        $-$ oracle-rewrite KD & 28.4 \\
        $-$ initialization anchor & 29.5 \\
        $-$ global negatives & 30.4 \\
        $-$ graded fine-tune & 28.6 \\
        $-$ LoRA (use full fine-tuning) & 32.2 \\
        $-$ low learning rate (use $1\mathrm{e}{-5}$) & 32.3 \\
        $-$ full-profile (use relevant-profile) & 32.0 \\
        \bottomrule
     \end{tabular}
     \label{tab:ablation}
\end{table}

\section{Submitted Runs}
\label{sec:runs}
The interactive task evaluates a complete conversational system online, through the iKAT user-simulation API (Sim.API): at each turn the system receives the simulated user's utterance, ranks passages from ClueWeb22-B, and returns a generated response. A retriever alone is therefore not enough, and, unlike the offline evaluation of Section~\ref{sec:experiments}, the profile is not given but accumulated on the fly. Each run wraps PCDR in a common per-turn pipeline: a gpt-5-mini model at low reasoning effort extracts new durable facts about the user from the latest exchange and adds them to the running profile that PCDR encodes; PCDR ranks passages, optionally fused with a second retriever by Reciprocal Rank Fusion (RRF)~\cite{cormack2009reciprocal} and optionally reranked; and the same model generates the response from the top five passages. The personalized dense component is always the PCDR model selected in Section~\ref{sec:hpsel}, which uses LoRA with $r{=}64$. The four runs differ only in this retrieval stack, placing PCDR in pipelines of increasing heterogeneity, from PCDR alone to fusion with retrievers progressively less similar to it. Table~\ref{tab:runs} summarizes them and gives their submission identifiers. Every run uses the PCDR dense leg and gpt-5-mini for profile extraction and response generation, and every fusion is RRF.

\paragraph{Run~6: PCDR alone.}
A single dense leg, with no reformulation, fusion, or reranking. The conversation and the profile are encoded by PCDR and passages are ranked by cosine similarity. This run measures the standalone effectiveness of personalized conversational dense retrieval.

\paragraph{Run~3: fusion with a conversational dense retriever.}
We fuse PCDR with its own non-personalized backbone, conv-Qwen3, by RRF and without reranking. Neither leg is reformulated, which makes the run fully rewrite-free: both encode the conversation directly, and only the personalized leg additionally consumes the profile. This design follows two observations from our 2024 system~\cite{hui2024rali}: fusing personalized and non-personalized rankings mitigates over-personalization, and a strong conversational dense retriever can replace the reformulate-then-retrieve pipeline.

\paragraph{Run~5: fusion with a learned sparse retriever.}
We fuse PCDR with SPLADE-v3~\cite{lassance2024splade}, a learned sparse retriever, by RRF and without reranking. SPLADE-v3 is driven by a de-contextualized, non-personalized reformulation of the current question, contributing profile-free lexical evidence that complements the personalized dense ranking. As in the preceding runs, we omit reranking here: when the fused pool is already strong, our 2024 experience~\cite{hui2024rali} indicates that a reranker brings little benefit and can even harm the ordering.

\paragraph{Run~4: fusion with a lexical retriever and reranking.}
We fuse PCDR with BM25~\cite{bm25} by RRF and rerank the top $50$ fused passages with RankLLaMA~\cite{ma2024rankllama}; BM25 is given a personalized reformulation of the question. As the lexical retriever is the weakest of the three fusion partners, we reinstate a reranking stage to consolidate the final ranking.

\begin{table}[t]
    \centering
    \caption{The four runs submitted to the iKAT 2026 interactive task. All runs use PCDR as the personalized dense retriever and gpt-5-mini for profile extraction and response generation, and all fusions use RRF.}
    \vspace{-2ex}
    \footnotesize
    \begin{tabular}{@{}cll@{}}
    \toprule
        Run & Retrieval components & Reranking \\
        \midrule
        6 & PCDR & --- \\
        3 & PCDR $+$ CDR & --- \\
        5 & PCDR $+$ SPLADE-v3 & --- \\
        4 & PCDR $+$ BM25 & RankLLaMA \\
        \bottomrule
     \end{tabular}

     \vspace{0.6ex}
     {\footnotesize\raggedright Submission identifiers. Run~6:
     \texttt{pers-\allowbreak conv-\allowbreak qwen-\allowbreak only}; Run~3:
     \texttt{conv-\allowbreak pers-\allowbreak qwen3-\allowbreak fuse-\allowbreak norerank-\allowbreak extract};
     Run~5: \texttt{splade-\allowbreak pers-\allowbreak conv-\allowbreak qwen}; Run~4:
     \texttt{pers-\allowbreak conv-\allowbreak dense-\allowbreak BM25-\allowbreak rankllama-\allowbreak gpt5mini-\allowbreak extract-\allowbreak new}.\par}
    \label{tab:runs}
\end{table}

\section{Results with Simulated Users}
\label{sec:simresults}

\subsection{Evaluation Setting}
The iKAT 2026 interactive task evaluates each submitted system online against LLM-based user simulators~\cite{abbasiantaeb2026ikat}. The evaluation reuses $15$ conversations drawn from the iKAT 2023, 2024, and 2025 test topics, five from each year, each with a distinct user profile. Each conversation is run against a subset of four user variants, $u_1$--$u_4$, yielding $44$ conversation--variant pairs: $u_4$ covers all $15$ conversations, while $u_1$, $u_2$, and $u_3$ cover $9$, $10$, and $10$ of them. For a given conversation, the variants share the \emph{same} information need and profile and differ only in simulated behaviour, for example in language proficiency or patience, so that a system is tested for robustness to how a user interacts rather than to what they want. The concrete behaviour of each variant is not disclosed by the organizers.

Both the simulation and the grading are driven by \emph{rubrics}: concise subtopic questions manually derived from each conversation, $92$ in total across the $15$ conversations, following the rubric-based evaluation methodology of Farzi and Dietz~\cite{farzi2024pencils}. At each turn the simulator produces an utterance for the current rubric; our system retrieves passages and generates a response; an LLM judge then scores the response against that rubric on a $0$--$5$ scale and advances to the next rubric once the score passes a threshold, otherwise letting the simulator give feedback and rephrase~\cite{abbasiantaeb2026ikat}. Three families of metrics are reported, all judged by an LLM; a manual evaluation is in progress at the time of writing.
\begin{itemize}
  \item \textbf{Retrieval.} The passages cited by our responses are pooled and judged against rubric-based graded qrels ($0$--$4$); we report turn-averaged NDCG@$5$.
  \item \textbf{Per-turn.} Each response is graded against the current rubric on four criteria: the \emph{rubric score}, which measures whether it answers the rubric question on a $0$--$5$ scale, and \emph{engagement}, \emph{relevance \& usefulness}, and \emph{overall subtopic quality} on a $1$--$5$ scale.
  \item \textbf{Dialogue-level.} The whole dialogue is rated on a $1$--$5$ scale for five criteria: \emph{personalization}, abbreviated Pers.; \emph{trustworthiness}, Trust; \emph{information flow \& coherence}, Flow; \emph{mixed-initiative strategies}, Mix-Init; and \emph{overall user satisfaction}, Sat.
\end{itemize}

\subsection{Run Comparison}
Tables~\ref{tab:sim-retrieval} and~\ref{tab:sim-dialogue} report the four submitted runs on the retrieval-plus-per-turn metrics and on the dialogue-level metrics, respectively. The \emph{Avg} column averages only the five-point criteria, namely the four rubric-level scores in Table~\ref{tab:sim-retrieval} and the five dialogue-level scores in Table~\ref{tab:sim-dialogue}. The retrieval NDCG@$5$, which ranges from $0$ to $1$, is kept separate and marked off by the dashed line. In both tables, higher is better.

\begin{table}[t]
    \centering
    \caption{Retrieval and per-turn response quality of the submitted runs with the iKAT 2026 simulated users, as judged by an LLM against the rubrics. NDCG@$5$ ranges from $0$ to $1$, the rubric score from $0$ to $5$, and the other criteria from $1$ to $5$; Avg averages the four response criteria.}
    \vspace{-2ex}
    \footnotesize
    \setlength{\tabcolsep}{4pt}
    \begin{tabular}{@{}lc:ccccc@{}}
    \toprule
        Run & NDCG@5 & Rubric & Eng. & Rel.\&Use. & Subtopic & Avg \\
        \midrule
        PCDR                 & 0.557 & 3.85 & 3.32 & 4.23 & 3.75 & 3.79 \\
        PCDR $+$ CDR  & 0.559 & 3.74 & 3.17 & 4.12 & 3.65 & 3.67 \\
        PCDR $+$ SPLADE-v3   & 0.577 & 3.97 & 3.20 & 4.11 & 3.69 & 3.74 \\
        PCDR $+$ BM25        & 0.580 & 3.93 & 3.14 & 4.24 & 3.79 & 3.78 \\
        \bottomrule
     \end{tabular}
    \label{tab:sim-retrieval}
\end{table}

\begin{table}[t]
    \centering
    \caption{Dialogue-level quality of the submitted runs with the iKAT 2026 simulated users, as judged by an LLM on a $1$--$5$ scale. Avg averages the five criteria.}
    \vspace{-2ex}
    \footnotesize
    \setlength{\tabcolsep}{4pt}
    \begin{tabular}{@{}lcccccc@{}}
    \toprule
        Run & Pers. & Trust & Flow & Mix-Init & Sat. & Avg \\
        \midrule
        PCDR                 & 3.93 & 3.68 & 3.98 & 3.00 & 3.93 & 3.70 \\
        PCDR $+$ CDR  & 3.82 & 3.70 & 3.98 & 2.70 & 4.02 & 3.64 \\
        PCDR $+$ SPLADE-v3   & 3.80 & 3.39 & 3.86 & 2.82 & 3.84 & 3.54 \\
        PCDR $+$ BM25        & 3.95 & 3.41 & 4.07 & 2.52 & 3.93 & 3.58 \\
        \bottomrule
     \end{tabular}
    \label{tab:sim-dialogue}
\end{table}

\subsection{Preliminary Analysis}
\noindent\textbf{Retrieval improves with more heterogeneous fusion.} PCDR alone already retrieves well, and fusion helps in proportion to how different the partner is from the dense leg. Fusing PCDR with its own non-personalized backbone conv-Qwen3, another dense retriever, barely moves the score; fusing with the learned-sparse SPLADE-v3 helps more, and fusing with lexical BM25 helps most. The ordering follows the complementarity of the evidence rather than the strength of the partner, since the added retrievers are individually weaker than PCDR. This is the chorus effect we observed in our 2024 system~\cite{hui2024rali}: rankings built from different matching signals cover different relevant passages, so a more heterogeneous fusion recovers more of them.

\noindent\textbf{An end-to-end, rewrite-free configuration.} The PCDR $+$ CDR run is fully rewrite-free: both legs are dense and encode the conversation and profile directly, so the entire pipeline consumes only the full conversation and the user profile, with no LLM query rewriting anywhere. It is the most genuinely end-to-end of the four runs and removes the latency and cost of an LLM rewriting step, and it remains competitive on all three families of metrics. This shows that the reformulate-then-retrieve stage can be dropped entirely without collapsing quality.

\noindent\textbf{Response quality is decoupled from retrieval.} The response-quality scores do not track the retrieval scores. PCDR alone has the lowest retrieval score yet obtains the highest response quality at both the per-turn and the dialogue level, while the fused runs with stronger retrieval score lower on both. Since every run generates its response from the same top passages with the same model, a better passage ranking does not translate into a better judged response here. We do not have an explanation for this inversion and leave it to future work.

\noindent\textbf{Mixed initiative.} Mixed-initiative strategies are our weakest dialogue dimension. This is expected, as our pipeline answers each question directly and we did not build an explicit mixed-initiative mechanism such as asking clarifying questions or proactively suggesting follow-ups~\cite{aliannejadi2021analysing}. The low scores reflect the absence of that mechanism rather than a retrieval deficiency.

\noindent\textbf{What we cannot yet conclude.} Three factors limit the analysis. First, other teams' submissions were not made available to us, so we can read only within-system contrasts and cannot place our results against the field. Second, a per-user-type analysis is not meaningful here, because the variants cover different and unequal subsets of the conversations, so averages per variant are confounded with topic difficulty; only six conversations are run with all four variants, which is too few for a controlled comparison. Third, the organizers do not disclose which user characteristics the four variants are meant to distinguish, which further prevents us from interpreting robustness across variants as intended.

\section{Visualization: What the Dense Encoder Attends To}
\label{sec:viz}
\begin{figure*}[t]
  \centering
  \includegraphics[width=\textwidth]{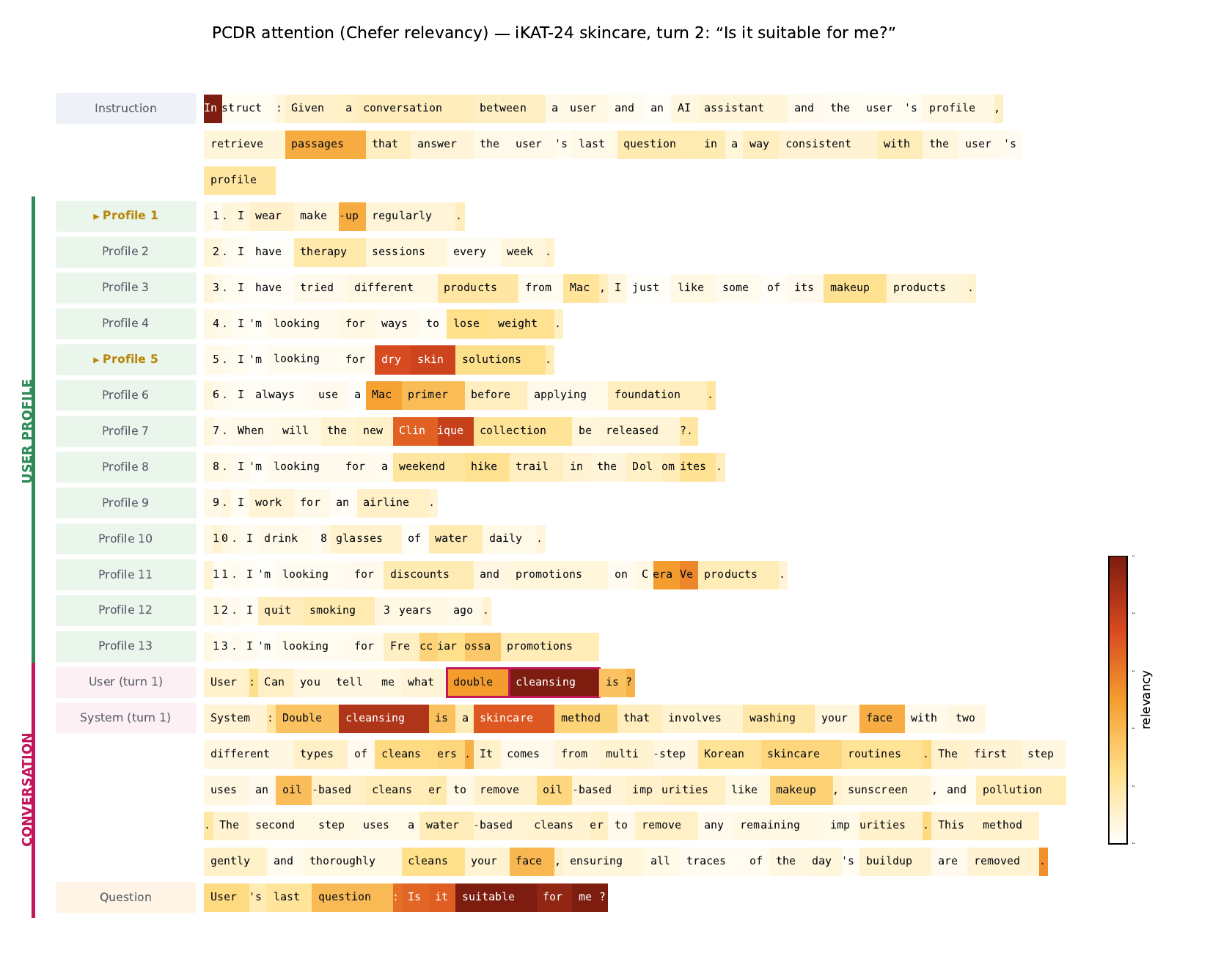}
  \vspace{-3ex}
  \caption{Token relevancy of the PCDR query embedding on an iKAT 2024 skincare turn, computed with the method of Chefer et al.~\cite{chefer2021generic}. Darker colors indicate a larger contribution to the embedding. The two profile statements annotated as relevant are marked with $\triangleright$, and the antecedent of ``it in the question is boxed.}
  \label{fig:chefer}
\end{figure*}

To make concrete the claim that PCDR denoises a long profile inside the representation, we visualize which input tokens most influence its query embedding. We use the generic attention-relevancy method of Chefer et al.~\cite{chefer2021generic}, which attributes a model output back to input tokens by aggregating attention maps weighted by their gradients across layers; a higher relevancy means a token contributes more to the retrieval embedding. Figure~\ref{fig:chefer} applies it to a representative personalized turn from iKAT 2024. The input consists of the instruction, a $13$-statement user profile, and a one-turn conversation ending in the question ``\textit{Is it suitable for me?}''. Only profile statements $1$, about makeup, and $5$, about dry skin, are annotated as relevant; they are marked with $\triangleright$ in the figure, and the antecedent of the pronoun is boxed.

The figure shows two behaviours that a dense encoder can perform but a bag-of-words query cannot. It attenuates the irrelevant part of the profile: of the thirteen statements, relevancy concentrates on the skin-related ones and on the current question, namely the discriminative tokens \textit{dry skin} and the topically related \textit{Mac primer}, \textit{Clinique}, and \textit{CeraVe}, while statements about travel, work, or water intake receive little. And it resolves conversational context, linking the pronoun ``\textit{it}'' in the question to ``\textit{double cleansing}'' introduced one turn earlier, without an explicit rewriting step.

The contrast with query reformulation is instructive. For the same turn, the personalized LLM rewrite our pipeline produces is
\begin{quote}\itshape
``Is double cleansing suitable for someone who wears makeup regularly and is looking for dry skin solutions?''
\end{quote}
To personalize, the rewrite must \emph{verbalize} the selected profile facts as additional query text, which lengthens the query and, when the profile is long, dilutes the discriminative terms among many others~\cite{Bendersky2008,Kumaran2009}. PCDR instead keeps the entire profile as input but internally attends to only the few relevant tokens shown in Figure~\ref{fig:chefer}, performing the same selection inside the representation rather than by editing text. The denoising happens in embedding space, which is exactly where dense retrieval is robust to verbosity and where reformulation is not.

\section{Conclusion}

In this paper, we tackle the problem of over-personalization caused by the excessive integration of user profile terms in existing personalized conversational query reformulation. We propose to perform user profile denoising in a latent semantic space, optimized directly for retrieval performance. In this way, more precise user information is integrated into the latent representation of the conversational information need to enhance search. Specifically, we train PCDR, which performs retrieval-aware user profile denoising directly in the dense retriever's embedding space. Initialized from a conversational dense retriever, PCDR requires only a small set of previous-year iKAT relevance judgments, augmented by light distillation from personalized human-rewrite embeddings. Online evaluation on the iKAT 2026 interactive runs shows that PCDR alone can produce satisfying tailored search results, and combining it with heterogeneous retrievers yields further improvements. Our visualization analysis further illustrates that the profile denoising goes beyond shallow semantic relevance.
Future work includes removing the reliance on distillation from human query rewrites, so that PCDR is trained from relevance signals alone and no longer needs rewritten queries at any stage. It also includes scaling training with synthesized end-to-end personalized conversational retrieval datasets and studying continual adaptation from ad-hoc search to conversational search and ultimately personalized conversational search.

\begin{acks}
We used generative AI tools for language editing and figure preparation; the authors checked all content and take full responsibility for it.
\end{acks}

\bibliographystyle{ACM-Reference-Format}
\bibliography{sample-base,software}


\begin{thebibliography}{66}

%%% the Web address.

\ifx \showCODEN    \undefined \def \showCODEN     #1{\unskip}     \fi
\ifx \showDOI      \undefined \def \showDOI       #1{#1}\fi
\ifx \showISBNx    \undefined \def \showISBNx     #1{\unskip}     \fi
\ifx \showISBNxiii \undefined \def \showISBNxiii  #1{\unskip}     \fi
\ifx \showISSN     \undefined \def \showISSN      #1{\unskip}     \fi
\ifx \showLCCN     \undefined \def \showLCCN      #1{\unskip}     \fi
\ifx \shownote     \undefined \def \shownote      #1{#1}          \fi
\ifx \showarticletitle \undefined \def \showarticletitle #1{#1}   \fi
\ifx \showURL      \undefined \def \showURL       {\relax}        \fi
% invisible to TeX
\providecommand\bibfield[2]{#2}
\providecommand\bibinfo[2]{#2}
\providecommand\natexlab[1]{#1}
\providecommand\showeprint[2][]{arXiv:#2}

\bibitem[Abbasiantaeb et~al\mbox{.}(2026)]%
        {abbasiantaeb2026ikat}
\bibfield{author}{\bibinfo{person}{Zahra Abbasiantaeb}, \bibinfo{person}{Simon
  Lupart}, \bibinfo{person}{Marcel Gohsen}, \bibinfo{person}{Nailia
  Mirzakhmedova}, \bibinfo{person}{Johannes Kiesel}, \bibinfo{person}{Jeffrey
  Dalton}, {and} \bibinfo{person}{Mohammad Aliannejadi}.}
  \bibinfo{year}{2026}\natexlab{}.
\newblock \showarticletitle{{TREC iKAT 2025}: A Test Collection for the Offline
  and Interactive Evaluation of Conversational Search}. In
  \bibinfo{booktitle}{\emph{Proceedings of the 49th International ACM SIGIR
  Conference on Research and Development in Information Retrieval (SIGIR)}}.
  \bibinfo{pages}{3092--3100}.
\newblock
\urldef\tempurl%
\url{https://doi.org/10.1145/3805712.3808591}
\showDOI{\tempurl}


\bibitem[Adlakha et~al\mbox{.}(2022)]%
        {adlakha2022topiocqa}
\bibfield{author}{\bibinfo{person}{Vaibhav Adlakha}, \bibinfo{person}{Shehzaad
  Dhuliawala}, \bibinfo{person}{Kaheer Suleman}, \bibinfo{person}{Harm de
  Vries}, {and} \bibinfo{person}{Siva Reddy}.} \bibinfo{year}{2022}\natexlab{}.
\newblock \showarticletitle{TopiOCQA: Open-domain Conversational Question
  Answering with Topic Switching}.
\newblock \bibinfo{journal}{\emph{Transactions of the Association for
  Computational Linguistics}}  \bibinfo{volume}{10} (\bibinfo{year}{2022}),
  \bibinfo{pages}{468--483}.
\newblock


\bibitem[Ai et~al\mbox{.}(2019)]%
        {ai2019zam}
\bibfield{author}{\bibinfo{person}{Qingyao Ai}, \bibinfo{person}{Daniel~N.
  Hill}, \bibinfo{person}{S.~V.~N. Vishwanathan}, {and}
  \bibinfo{person}{W.~Bruce Croft}.} \bibinfo{year}{2019}\natexlab{}.
\newblock \showarticletitle{A Zero Attention Model for Personalized Product
  Search}. In \bibinfo{booktitle}{\emph{Proceedings of the 28th ACM
  International Conference on Information and Knowledge Management (CIKM)}}.
  \bibinfo{pages}{379--388}.
\newblock
\urldef\tempurl%
\url{https://doi.org/10.1145/3357384.3357980}
\showDOI{\tempurl}


\bibitem[Ai et~al\mbox{.}(2017)]%
        {ai2017hem}
\bibfield{author}{\bibinfo{person}{Qingyao Ai}, \bibinfo{person}{Yongfeng
  Zhang}, \bibinfo{person}{Keping Bi}, \bibinfo{person}{Xu Chen}, {and}
  \bibinfo{person}{W.~Bruce Croft}.} \bibinfo{year}{2017}\natexlab{}.
\newblock \showarticletitle{Learning a Hierarchical Embedding Model for
  Personalized Product Search}. In \bibinfo{booktitle}{\emph{Proceedings of the
  40th International ACM SIGIR Conference on Research and Development in
  Information Retrieval}}. \bibinfo{pages}{645--654}.
\newblock
\urldef\tempurl%
\url{https://doi.org/10.1145/3077136.3080813}
\showDOI{\tempurl}


\bibitem[Aliannejadi et~al\mbox{.}(2024)]%
        {aliannejadi2024ikat}
\bibfield{author}{\bibinfo{person}{Mohammad Aliannejadi},
  \bibinfo{person}{Zahra Abbasiantaeb}, \bibinfo{person}{Shubham Chatterjee},
  \bibinfo{person}{Jeffrey Dalton}, {and} \bibinfo{person}{Leif Azzopardi}.}
  \bibinfo{year}{2024}\natexlab{}.
\newblock \showarticletitle{{TREC iKAT 2023}: A Test Collection for Evaluating
  Conversational and Interactive Knowledge Assistants}. In
  \bibinfo{booktitle}{\emph{Proceedings of the 47th International ACM SIGIR
  Conference on Research and Development in Information Retrieval}}.
  \bibinfo{pages}{819--829}.
\newblock
\urldef\tempurl%
\url{https://doi.org/10.1145/3626772.3657860}
\showDOI{\tempurl}


\bibitem[Aliannejadi et~al\mbox{.}(2021)]%
        {aliannejadi2021analysing}
\bibfield{author}{\bibinfo{person}{Mohammad Aliannejadi}, \bibinfo{person}{Leif
  Azzopardi}, \bibinfo{person}{Hamed Zamani}, \bibinfo{person}{Evangelos
  Kanoulas}, \bibinfo{person}{Paul Thomas}, {and} \bibinfo{person}{Nick
  Craswell}.} \bibinfo{year}{2021}\natexlab{}.
\newblock \showarticletitle{Analysing mixed initiatives and search strategies
  during conversational search}. In \bibinfo{booktitle}{\emph{Proceedings of
  the 30th ACM International Conference on Information \& Knowledge
  Management}}. \bibinfo{pages}{16--26}.
\newblock


\bibitem[Bao et~al\mbox{.}(2024)]%
        {bao2024pr2}
\bibfield{author}{\bibinfo{person}{Wentian Bao}, \bibinfo{person}{Hu Liu},
  \bibinfo{person}{Kai Zheng}, \bibinfo{person}{Chao Zhang},
  \bibinfo{person}{Shunyu Zhang}, \bibinfo{person}{Enyun Yu},
  \bibinfo{person}{Wenwu Ou}, {and} \bibinfo{person}{Yang Song}.}
  \bibinfo{year}{2024}\natexlab{}.
\newblock \showarticletitle{Beyond Relevance: Improving User Engagement by
  Personalization for Short-Video Search}.
\newblock \bibinfo{journal}{\emph{arXiv preprint arXiv:2409.11281}}
  (\bibinfo{year}{2024}).
\newblock


\bibitem[Belyi et~al\mbox{.}(2023)]%
        {belyi2023personalized}
\bibfield{author}{\bibinfo{person}{Masha Belyi}, \bibinfo{person}{Charlotte
  Dzialo}, \bibinfo{person}{Chaitanya Dwivedi}, \bibinfo{person}{Prajit
  Muppidi}, {and} \bibinfo{person}{Kanna Shimizu}.}
  \bibinfo{year}{2023}\natexlab{}.
\newblock \showarticletitle{Personalized Dense Retrieval on Global Index for
  Voice-enabled Conversational Systems}. In
  \bibinfo{booktitle}{\emph{Proceedings of the 2023 Conference on Empirical
  Methods in Natural Language Processing: Industry Track}}.
  \bibinfo{pages}{83--92}.
\newblock
\urldef\tempurl%
\url{https://doi.org/10.18653/v1/2023.emnlp-industry.9}
\showDOI{\tempurl}


\bibitem[Bendersky and Croft(2008)]%
        {Bendersky2008}
\bibfield{author}{\bibinfo{person}{Michael Bendersky} {and}
  \bibinfo{person}{W.~Bruce Croft}.} \bibinfo{year}{2008}\natexlab{}.
\newblock \showarticletitle{Discovering Key Concepts in Verbose Queries}. In
  \bibinfo{booktitle}{\emph{Proceedings of the 31st Annual International ACM
  SIGIR Conference on Research and Development in Information Retrieval (SIGIR
  '08)}}. \bibinfo{pages}{491--498}.
\newblock


\bibitem[Bi et~al\mbox{.}(2020)]%
        {bi2020tem}
\bibfield{author}{\bibinfo{person}{Keping Bi}, \bibinfo{person}{Qingyao Ai},
  {and} \bibinfo{person}{W.~Bruce Croft}.} \bibinfo{year}{2020}\natexlab{}.
\newblock \showarticletitle{A Transformer-based Embedding Model for
  Personalized Product Search}. In \bibinfo{booktitle}{\emph{Proceedings of the
  43rd International ACM SIGIR Conference on Research and Development in
  Information Retrieval}}. \bibinfo{pages}{1521--1524}.
\newblock
\urldef\tempurl%
\url{https://doi.org/10.1145/3397271.3401192}
\showDOI{\tempurl}


\bibitem[Chefer et~al\mbox{.}(2021)]%
        {chefer2021generic}
\bibfield{author}{\bibinfo{person}{Hila Chefer}, \bibinfo{person}{Shir Gur},
  {and} \bibinfo{person}{Lior Wolf}.} \bibinfo{year}{2021}\natexlab{}.
\newblock \showarticletitle{Generic Attention-model Explainability for
  Interpreting Bi-Modal and Encoder-Decoder Transformers}. In
  \bibinfo{booktitle}{\emph{Proceedings of the IEEE/CVF International
  Conference on Computer Vision (ICCV)}}. \bibinfo{pages}{397--406}.
\newblock


\bibitem[Chen et~al\mbox{.}(2024)]%
        {chen2024generalizing}
\bibfield{author}{\bibinfo{person}{Haonan Chen}, \bibinfo{person}{Zhicheng
  Dou}, \bibinfo{person}{Kelong Mao}, \bibinfo{person}{Jiongnan Liu}, {and}
  \bibinfo{person}{Ziliang Zhao}.} \bibinfo{year}{2024}\natexlab{}.
\newblock \showarticletitle{Generalizing Conversational Dense Retrieval via
  LLM-Cognition Data Augmentation}.
\newblock \bibinfo{journal}{\emph{arXiv preprint arXiv:2402.07092}}
  (\bibinfo{year}{2024}).
\newblock


\bibitem[Cormack et~al\mbox{.}(2009)]%
        {cormack2009reciprocal}
\bibfield{author}{\bibinfo{person}{Gordon~V Cormack}, \bibinfo{person}{Charles
  L~A Clarke}, {and} \bibinfo{person}{Stefan Buettcher}.}
  \bibinfo{year}{2009}\natexlab{}.
\newblock \showarticletitle{Reciprocal Rank Fusion Outperforms {Condorcet} and
  Individual Rank Learning Methods}. In \bibinfo{booktitle}{\emph{Proceedings
  of the 32nd International ACM SIGIR Conference on Research and Development in
  Information Retrieval}}.
\newblock


\bibitem[Dong et~al\mbox{.}(2026)]%
        {dong2026apr}
\bibfield{author}{\bibinfo{person}{Shen Dong}, \bibinfo{person}{Iadh Ounis},
  {and} \bibinfo{person}{Debasis Ganguly}.} \bibinfo{year}{2026}\natexlab{}.
\newblock \showarticletitle{{APR}: Adaptive Personalised Reranking for
  Conversational Search}. In \bibinfo{booktitle}{\emph{Proceedings of the 49th
  International ACM SIGIR Conference on Research and Development in Information
  Retrieval}}. \bibinfo{pages}{3694--3699}.
\newblock
\urldef\tempurl%
\url{https://doi.org/10.1145/3805712.3809914}
\showDOI{\tempurl}


\bibitem[Dong et~al\mbox{.}(2025)]%
        {dong2025personalise}
\bibfield{author}{\bibinfo{person}{Shen Dong}, \bibinfo{person}{Iadh Ounis},
  {and} \bibinfo{person}{Zaiqiao Meng}.} \bibinfo{year}{2025}\natexlab{}.
\newblock \showarticletitle{To Personalise or Not? {W}hy Personalisation
  Knowledge Can Fall Short in Enhancing Conversational Information Seeking}. In
  \bibinfo{booktitle}{\emph{Proceedings of the Second Workshop on
  Knowledge-Enhanced Information Retrieval (KEIR @ ECIR 2025)}}.
\newblock


\bibitem[Farzi and Dietz(2024)]%
        {farzi2024pencils}
\bibfield{author}{\bibinfo{person}{Naghmeh Farzi} {and} \bibinfo{person}{Laura
  Dietz}.} \bibinfo{year}{2024}\natexlab{}.
\newblock \showarticletitle{Pencils Down! Automatic Rubric-based Evaluation of
  Retrieve/Generate Systems}. In \bibinfo{booktitle}{\emph{Proceedings of the
  2024 ACM SIGIR International Conference on Theory of Information Retrieval
  (ICTIR)}}. \bibinfo{pages}{175--184}.
\newblock


\bibitem[Gao et~al\mbox{.}(2022)]%
        {gao2022neural}
\bibfield{author}{\bibinfo{person}{Jianfeng Gao}, \bibinfo{person}{Chenyan
  Xiong}, \bibinfo{person}{Paul Bennett}, {and} \bibinfo{person}{Nick
  Craswell}.} \bibinfo{year}{2022}\natexlab{}.
\newblock \showarticletitle{Neural approaches to conversational information
  retrieval}.
\newblock \bibinfo{journal}{\emph{arXiv preprint arXiv:2201.05176}}
  (\bibinfo{year}{2022}).
\newblock


\bibitem[Hu et~al\mbox{.}(2022)]%
        {hu2022lora}
\bibfield{author}{\bibinfo{person}{Edward~J Hu}, \bibinfo{person}{Yelong Shen},
  \bibinfo{person}{Phillip Wallis}, \bibinfo{person}{Zeyuan Allen-Zhu},
  \bibinfo{person}{Yuanzhi Li}, \bibinfo{person}{Shean Wang},
  \bibinfo{person}{Lu Wang}, {and} \bibinfo{person}{Weizhu Chen}.}
  \bibinfo{year}{2022}\natexlab{}.
\newblock \showarticletitle{{LoRA}: Low-Rank Adaptation of Large Language
  Models}. In \bibinfo{booktitle}{\emph{International Conference on Learning
  Representations (ICLR)}}.
\newblock


\bibitem[Huang et~al\mbox{.}(2020)]%
        {huang2020ebr}
\bibfield{author}{\bibinfo{person}{Jui-Ting Huang}, \bibinfo{person}{Ashish
  Sharma}, \bibinfo{person}{Shuying Sun}, \bibinfo{person}{Li Xia},
  \bibinfo{person}{David Zhang}, \bibinfo{person}{Philip Pronin},
  \bibinfo{person}{Janani Padmanabhan}, \bibinfo{person}{Giuseppe Ottaviano},
  {and} \bibinfo{person}{Linjun Yang}.} \bibinfo{year}{2020}\natexlab{}.
\newblock \showarticletitle{Embedding-based Retrieval in {Facebook} Search}. In
  \bibinfo{booktitle}{\emph{Proceedings of the 26th ACM SIGKDD Conference on
  Knowledge Discovery and Data Mining}}. \bibinfo{pages}{2553--2561}.
\newblock
\urldef\tempurl%
\url{https://doi.org/10.1145/3394486.3403305}
\showDOI{\tempurl}


\bibitem[Hui et~al\mbox{.}(2024)]%
        {hui2024rali}
\bibfield{author}{\bibinfo{person}{Yuchen Hui}, \bibinfo{person}{Fengran Mo},
  \bibinfo{person}{Kelong Mao}, {and} \bibinfo{person}{Jian-Yun Nie}.}
  \bibinfo{year}{2024}\natexlab{}.
\newblock \showarticletitle{{RALI@TREC iKAT 2024}: Achieving Personalization
  via Retrieval Fusion in Conversational Search}.
\newblock \bibinfo{journal}{\emph{arXiv preprint arXiv:2412.07998}}
  (\bibinfo{year}{2024}).
\newblock


\bibitem[Jha et~al\mbox{.}(2024)]%
        {jha2024etsy}
\bibfield{author}{\bibinfo{person}{Rishikesh Jha}, \bibinfo{person}{Siddharth
  Subramaniyam}, \bibinfo{person}{Ethan Benjamin}, {and}
  \bibinfo{person}{Thrivikrama Taula}.} \bibinfo{year}{2024}\natexlab{}.
\newblock \showarticletitle{Unified Embedding Based Personalized Retrieval in
  {Etsy} Search}. In \bibinfo{booktitle}{\emph{2024 IEEE International
  Conference on Future Machine Learning and Data Science (FMLDS)}}.
  \bibinfo{pages}{258--264}.
\newblock
\urldef\tempurl%
\url{https://doi.org/10.1109/FMLDS63805.2024.00055}
\showDOI{\tempurl}


\bibitem[Jiang et~al\mbox{.}(2026)]%
        {jiang2026ttp}
\bibfield{author}{\bibinfo{person}{Angqing Jiang}, \bibinfo{person}{Gaoming
  Zhang}, \bibinfo{person}{Jianchun Song}, \bibinfo{person}{Kena Qi},
  \bibinfo{person}{Dayao Chen}, \bibinfo{person}{Wei Lin}, {and}
  \bibinfo{person}{Defu Lian}.} \bibinfo{year}{2026}\natexlab{}.
\newblock \showarticletitle{Think-to-Personalize: Unifying Reasoning and
  Retrieval for User-Centric Personalized Dense Retrieval}.
\newblock \bibinfo{journal}{\emph{arXiv preprint arXiv:2608.18855}}
  (\bibinfo{year}{2026}).
\newblock
\newblock
\shownote{Accepted at CIKM 2026}.


\bibitem[Jiang(2026)]%
        {jiang2026onemodel}
\bibfield{author}{\bibinfo{person}{Bo Jiang}.} \bibinfo{year}{2026}\natexlab{}.
\newblock \showarticletitle{One Model Is Enough: Native Retrieval Embeddings
  from {LLM} Agent Hidden States}.
\newblock \bibinfo{journal}{\emph{arXiv preprint arXiv:2603.08429}}
  (\bibinfo{year}{2026}).
\newblock


\bibitem[Jin et~al\mbox{.}(2023)]%
        {jin2023instructor}
\bibfield{author}{\bibinfo{person}{Zhuoran Jin}, \bibinfo{person}{Pengfei Cao},
  \bibinfo{person}{Yubo Chen}, \bibinfo{person}{Kang Liu}, {and}
  \bibinfo{person}{Jun Zhao}.} \bibinfo{year}{2023}\natexlab{}.
\newblock \showarticletitle{InstructoR: Instructing Unsupervised Conversational
  Dense Retrieval with Large Language Models}. In
  \bibinfo{booktitle}{\emph{Findings of the Association for Computational
  Linguistics: EMNLP 2023}}. \bibinfo{pages}{6649--6675}.
\newblock


\bibitem[Kim and Kim(2022)]%
        {kim2022saving}
\bibfield{author}{\bibinfo{person}{Sungdong Kim} {and} \bibinfo{person}{Gangwoo
  Kim}.} \bibinfo{year}{2022}\natexlab{}.
\newblock \showarticletitle{Saving dense retriever from shortcut dependency in
  conversational search}. In \bibinfo{booktitle}{\emph{Proceedings of the 2022
  Conference on Empirical Methods in Natural Language Processing}}.
  \bibinfo{publisher}{Association for Computational Linguistics},
  \bibinfo{pages}{10278--10287}.
\newblock


\bibitem[Kumaran and Carvalho(2009)]%
        {Kumaran2009}
\bibfield{author}{\bibinfo{person}{Giridhar Kumaran} {and}
  \bibinfo{person}{Vitor~R. Carvalho}.} \bibinfo{year}{2009}\natexlab{}.
\newblock \showarticletitle{Reducing Long Queries Using Query Quality
  Predictors}. In \bibinfo{booktitle}{\emph{Proceedings of the 32nd
  International ACM SIGIR Conference on Research and Development in Information
  Retrieval (SIGIR '09)}}. \bibinfo{pages}{564--571}.
\newblock


\bibitem[Lassance et~al\mbox{.}(2024)]%
        {lassance2024splade}
\bibfield{author}{\bibinfo{person}{Carlos Lassance}, \bibinfo{person}{Herv{\'e}
  D{\'e}jean}, \bibinfo{person}{Thibault Formal}, {and}
  \bibinfo{person}{St{\'e}phane Clinchant}.} \bibinfo{year}{2024}\natexlab{}.
\newblock \showarticletitle{{SPLADE-v3}: New baselines for {SPLADE}}.
\newblock \bibinfo{journal}{\emph{arXiv preprint arXiv:2403.06789}}
  (\bibinfo{year}{2024}).
\newblock


\bibitem[Li et~al\mbox{.}(2021)]%
        {li2021mgdspr}
\bibfield{author}{\bibinfo{person}{Sen Li}, \bibinfo{person}{Fuyu Lv},
  \bibinfo{person}{Taiwei Jin}, \bibinfo{person}{Guli Lin},
  \bibinfo{person}{Keping Yang}, \bibinfo{person}{Xiaoyi Zeng},
  \bibinfo{person}{Xiao-Ming Wu}, {and} \bibinfo{person}{Qianli Ma}.}
  \bibinfo{year}{2021}\natexlab{}.
\newblock \showarticletitle{Embedding-based Product Retrieval in {Taobao}
  Search}. In \bibinfo{booktitle}{\emph{Proceedings of the 27th ACM SIGKDD
  Conference on Knowledge Discovery and Data Mining}}.
  \bibinfo{pages}{3181--3189}.
\newblock
\urldef\tempurl%
\url{https://doi.org/10.1145/3447548.3467101}
\showDOI{\tempurl}


\bibitem[Lin et~al\mbox{.}(2021)]%
        {lin2021contextualized}
\bibfield{author}{\bibinfo{person}{Sheng-Chieh Lin},
  \bibinfo{person}{Jheng-Hong Yang}, {and} \bibinfo{person}{Jimmy Lin}.}
  \bibinfo{year}{2021}\natexlab{}.
\newblock \showarticletitle{Contextualized Query Embeddings for Conversational
  Search}. In \bibinfo{booktitle}{\emph{Proceedings of the 2021 Conference on
  Empirical Methods in Natural Language Processing}}.
  \bibinfo{pages}{1004--1015}.
\newblock


\bibitem[Lupart et~al\mbox{.}(2024)]%
        {lupart2024irlab}
\bibfield{author}{\bibinfo{person}{Simon Lupart}, \bibinfo{person}{Zahra
  Abbasiantaeb}, {and} \bibinfo{person}{Mohammad Aliannejadi}.}
  \bibinfo{year}{2024}\natexlab{}.
\newblock \showarticletitle{{IRLab@iKAT24}: Learned Sparse Retrieval with
  Multi-aspect {LLM} Query Generation for Conversational Search}. In
  \bibinfo{booktitle}{\emph{Proceedings of the Thirty-Third Text REtrieval
  Conference (TREC 2024)}}.
\newblock


\bibitem[Lupart et~al\mbox{.}(2025)]%
        {lupart2025disco}
\bibfield{author}{\bibinfo{person}{Simon Lupart}, \bibinfo{person}{Mohammad
  Aliannejadi}, {and} \bibinfo{person}{Evangelos Kanoulas}.}
  \bibinfo{year}{2025}\natexlab{}.
\newblock \showarticletitle{{DiSCo}: {LLM} Knowledge Distillation for Efficient
  Sparse Retrieval in Conversational Search}. In
  \bibinfo{booktitle}{\emph{Proceedings of the 48th International ACM SIGIR
  Conference on Research and Development in Information Retrieval}}.
  \bibinfo{pages}{9--19}.
\newblock
\urldef\tempurl%
\url{https://doi.org/10.1145/3726302.3729966}
\showDOI{\tempurl}


\bibitem[Ma et~al\mbox{.}(2024)]%
        {ma2024rankllama}
\bibfield{author}{\bibinfo{person}{Xueguang Ma}, \bibinfo{person}{Liang Wang},
  \bibinfo{person}{Nan Yang}, \bibinfo{person}{Furu Wei}, {and}
  \bibinfo{person}{Jimmy Lin}.} \bibinfo{year}{2024}\natexlab{}.
\newblock \showarticletitle{Fine-Tuning {LLaMA} for Multi-Stage Text
  Retrieval}. In \bibinfo{booktitle}{\emph{Proceedings of the 47th
  International ACM SIGIR Conference on Research and Development in Information
  Retrieval}}.
\newblock


\bibitem[Mao et~al\mbox{.}(2024)]%
        {mao2024chatretriever}
\bibfield{author}{\bibinfo{person}{Kelong Mao}, \bibinfo{person}{Chenlong
  Deng}, \bibinfo{person}{Haonan Chen}, \bibinfo{person}{Fengran Mo},
  \bibinfo{person}{Zheng Liu}, \bibinfo{person}{Tetsuya Sakai}, {and}
  \bibinfo{person}{Zhicheng Dou}.} \bibinfo{year}{2024}\natexlab{}.
\newblock \showarticletitle{ChatRetriever: Adapting Large Language Models for
  Generalized and Robust Conversational Dense Retrieval}.
\newblock \bibinfo{journal}{\emph{arXiv preprint arXiv:2404.13556}}
  (\bibinfo{year}{2024}).
\newblock


\bibitem[Mao et~al\mbox{.}(2023a)]%
        {mao2023large}
\bibfield{author}{\bibinfo{person}{Kelong Mao}, \bibinfo{person}{Zhicheng Dou},
  \bibinfo{person}{Haonan Chen}, \bibinfo{person}{Fengran Mo}, {and}
  \bibinfo{person}{Hongjin Qian}.} \bibinfo{year}{2023}\natexlab{a}.
\newblock \showarticletitle{Large Language Models Know Your Contextual Search
  Intent: A Prompting Framework for Conversational Search}. In
  \bibinfo{booktitle}{\emph{Findings of the Association for Computational
  Linguistics: EMNLP 2023}}.
\newblock


\bibitem[Mao et~al\mbox{.}(2022a)]%
        {mao2022curriculum}
\bibfield{author}{\bibinfo{person}{Kelong Mao}, \bibinfo{person}{Zhicheng Dou},
  {and} \bibinfo{person}{Hongjin Qian}.} \bibinfo{year}{2022}\natexlab{a}.
\newblock \showarticletitle{Curriculum Contrastive Context Denoising for
  Few-shot Conversational Dense Retrieval}. In
  \bibinfo{booktitle}{\emph{Proceedings of the 45th International ACM SIGIR
  Conference on Research and Development in Information Retrieval}}.
  \bibinfo{pages}{176--186}.
\newblock


\bibitem[Mao et~al\mbox{.}(2022b)]%
        {mao2022convtrans}
\bibfield{author}{\bibinfo{person}{Kelong Mao}, \bibinfo{person}{Zhicheng Dou},
  \bibinfo{person}{Hongjin Qian}, \bibinfo{person}{Fengran Mo},
  \bibinfo{person}{Xiaohua Cheng}, {and} \bibinfo{person}{Zhao Cao}.}
  \bibinfo{year}{2022}\natexlab{b}.
\newblock \showarticletitle{ConvTrans: Transforming Web Search Sessions for
  Conversational Dense Retrieval}. In \bibinfo{booktitle}{\emph{Proceedings of
  the 2022 Conference on Empirical Methods in Natural Language Processing}}.
  \bibinfo{pages}{2935--2946}.
\newblock


\bibitem[Mao et~al\mbox{.}(2023b)]%
        {mao2023learning}
\bibfield{author}{\bibinfo{person}{Kelong Mao}, \bibinfo{person}{Hongjin Qian},
  \bibinfo{person}{Fengran Mo}, \bibinfo{person}{Zhicheng Dou},
  \bibinfo{person}{Bang Liu}, \bibinfo{person}{Xiaohua Cheng}, {and}
  \bibinfo{person}{Zhao Cao}.} \bibinfo{year}{2023}\natexlab{b}.
\newblock \showarticletitle{Learning Denoised and Interpretable Session
  Representation for Conversational Search}. In
  \bibinfo{booktitle}{\emph{Proceedings of the ACM Web Conference 2023}}.
  \bibinfo{pages}{3193--3202}.
\newblock


\bibitem[Mo et~al\mbox{.}(2025a)]%
        {mo2025uniconv}
\bibfield{author}{\bibinfo{person}{Fengran Mo}, \bibinfo{person}{Yifan Gao},
  \bibinfo{person}{Chuan Meng}, \bibinfo{person}{Xin Liu},
  \bibinfo{person}{Zhuofeng Wu}, \bibinfo{person}{Kelong Mao},
  \bibinfo{person}{Zhengyang Wang}, \bibinfo{person}{Pei Chen},
  \bibinfo{person}{Zheng Li}, \bibinfo{person}{Xian Li}, \bibinfo{person}{Bing
  Yin}, {and} \bibinfo{person}{Meng Jiang}.} \bibinfo{year}{2025}\natexlab{a}.
\newblock \showarticletitle{{UniConv}: Unifying Retrieval and Response
  Generation for Large Language Models in Conversations}. In
  \bibinfo{booktitle}{\emph{Proceedings of the 63rd Annual Meeting of the
  Association for Computational Linguistics (ACL)}}.
  \bibinfo{pages}{6936--6949}.
\newblock
\urldef\tempurl%
\url{https://doi.org/10.18653/v1/2025.acl-long.344}
\showDOI{\tempurl}


\bibitem[Mo et~al\mbox{.}(2024a)]%
        {mo2024chiq}
\bibfield{author}{\bibinfo{person}{Fengran Mo}, \bibinfo{person}{Abbas
  Ghaddar}, \bibinfo{person}{Kelong Mao}, \bibinfo{person}{Mehdi
  Rezagholizadeh}, \bibinfo{person}{Boxing Chen}, \bibinfo{person}{Qun Liu},
  {and} \bibinfo{person}{Jian-Yun Nie}.} \bibinfo{year}{2024}\natexlab{a}.
\newblock \showarticletitle{CHIQ: Contextual History Enhancement for Improving
  Query Rewriting in Conversational Search}.
\newblock \bibinfo{journal}{\emph{arXiv preprint arXiv:2406.05013}}
  (\bibinfo{year}{2024}).
\newblock


\bibitem[Mo et~al\mbox{.}(2025b)]%
        {mo2025apcir}
\bibfield{author}{\bibinfo{person}{Fengran Mo}, \bibinfo{person}{Yuchen Hui},
  \bibinfo{person}{Yuxing Tian}, \bibinfo{person}{Zhaoxuan Tan},
  \bibinfo{person}{Chuan Meng}, \bibinfo{person}{Zhan Su},
  \bibinfo{person}{Kaiyu Huang}, {and} \bibinfo{person}{Jian-Yun Nie}.}
  \bibinfo{year}{2025}\natexlab{b}.
\newblock \showarticletitle{Towards Adaptive Personalized Conversational
  Information Retrieval}. In \bibinfo{booktitle}{\emph{Proceedings of the 34th
  ACM International Conference on Information and Knowledge Management
  (CIKM)}}. \bibinfo{pages}{2137--2147}.
\newblock
\urldef\tempurl%
\url{https://doi.org/10.1145/3746252.3761255}
\showDOI{\tempurl}


\bibitem[Mo et~al\mbox{.}(2025c)]%
        {mo2025survey}
\bibfield{author}{\bibinfo{person}{Fengran Mo}, \bibinfo{person}{Kelong Mao},
  \bibinfo{person}{Ziliang Zhao}, \bibinfo{person}{Hongjin Qian},
  \bibinfo{person}{Haonan Chen}, \bibinfo{person}{Yiruo Cheng},
  \bibinfo{person}{Xiaoxi Li}, \bibinfo{person}{Yutao Zhu},
  \bibinfo{person}{Zhicheng Dou}, {and} \bibinfo{person}{Jian-Yun Nie}.}
  \bibinfo{year}{2025}\natexlab{c}.
\newblock \showarticletitle{A Survey of Conversational Search}.
\newblock \bibinfo{journal}{\emph{ACM Transactions on Information Systems}}
  \bibinfo{volume}{43}, \bibinfo{number}{6} (\bibinfo{year}{2025}),
  \bibinfo{pages}{1--50}.
\newblock
\urldef\tempurl%
\url{https://doi.org/10.1145/3759453}
\showDOI{\tempurl}


\bibitem[Mo et~al\mbox{.}(2023a)]%
        {mo2023convgqr}
\bibfield{author}{\bibinfo{person}{Fengran Mo}, \bibinfo{person}{Kelong Mao},
  \bibinfo{person}{Yutao Zhu}, \bibinfo{person}{Yihong Wu},
  \bibinfo{person}{Kaiyu Huang}, {and} \bibinfo{person}{Jian-Yun Nie}.}
  \bibinfo{year}{2023}\natexlab{a}.
\newblock \showarticletitle{ConvGQR: Generative Query Reformulation for
  Conversational Search}. In \bibinfo{booktitle}{\emph{Proceedings of the 61st
  Annual Meeting of the Association for Computational Linguistics}}.
  \bibinfo{pages}{4998–5012}.
\newblock


\bibitem[Mo et~al\mbox{.}(2023b)]%
        {mo2023learning}
\bibfield{author}{\bibinfo{person}{Fengran Mo}, \bibinfo{person}{Jian-Yun Nie},
  \bibinfo{person}{Kaiyu Huang}, \bibinfo{person}{Kelong Mao},
  \bibinfo{person}{Yutao Zhu}, \bibinfo{person}{Peng Li}, {and}
  \bibinfo{person}{Yang Liu}.} \bibinfo{year}{2023}\natexlab{b}.
\newblock \showarticletitle{Learning to Relate to Previous Turns in
  Conversational Search}. In \bibinfo{booktitle}{\emph{29th ACM SIGKDD
  Conference On Knowledge Discover and Data Mining (SIGKDD)}}.
\newblock


\bibitem[Mo et~al\mbox{.}(2024b)]%
        {mo2024aligning}
\bibfield{author}{\bibinfo{person}{Fengran Mo}, \bibinfo{person}{Chen Qu},
  \bibinfo{person}{Kelong Mao}, \bibinfo{person}{Yihong Wu},
  \bibinfo{person}{Zhan Su}, \bibinfo{person}{Kaiyu Huang}, {and}
  \bibinfo{person}{Jian-Yun Nie}.} \bibinfo{year}{2024}\natexlab{b}.
\newblock \showarticletitle{Aligning Query Representation with Rewritten Query
  and Relevance Judgments in Conversational Search}. In
  \bibinfo{booktitle}{\emph{Proceedings of the 33rd ACM International
  Conference on Information and Knowledge Management (CIKM)}}.
  \bibinfo{pages}{1700--1710}.
\newblock
\urldef\tempurl%
\url{https://doi.org/10.1145/3627673.3679534}
\showDOI{\tempurl}


\bibitem[Mo et~al\mbox{.}(2024c)]%
        {mo2024history}
\bibfield{author}{\bibinfo{person}{Fengran Mo}, \bibinfo{person}{Chen Qu},
  \bibinfo{person}{Kelong Mao}, \bibinfo{person}{Tianyu Zhu},
  \bibinfo{person}{Zhan Su}, \bibinfo{person}{Kaiyu Huang}, {and}
  \bibinfo{person}{Jian-Yun Nie}.} \bibinfo{year}{2024}\natexlab{c}.
\newblock \showarticletitle{History-Aware Conversational Dense Retrieval}.
\newblock \bibinfo{journal}{\emph{arXiv preprint arXiv:2401.16659}}
  (\bibinfo{year}{2024}).
\newblock


\bibitem[Mo et~al\mbox{.}(2024d)]%
        {mo2024convsdg}
\bibfield{author}{\bibinfo{person}{Fengran Mo}, \bibinfo{person}{Bole Yi},
  \bibinfo{person}{Kelong Mao}, \bibinfo{person}{Chen Qu},
  \bibinfo{person}{Kaiyu Huang}, {and} \bibinfo{person}{Jian-Yun Nie}.}
  \bibinfo{year}{2024}\natexlab{d}.
\newblock \showarticletitle{ConvSDG: Session Data Generation for Conversational
  Search}. In \bibinfo{booktitle}{\emph{Companion Proceedings of the ACM on Web
  Conference 2024}}. \bibinfo{pages}{1634--1642}.
\newblock


\bibitem[Mo et~al\mbox{.}(2026)]%
        {mo2026convmix}
\bibfield{author}{\bibinfo{person}{Fengran Mo}, \bibinfo{person}{Jinghan
  Zhang}, \bibinfo{person}{Yuchen Hui}, \bibinfo{person}{Jia~Ao Sun},
  \bibinfo{person}{Zhichao Xu}, \bibinfo{person}{Zhan Su}, {and}
  \bibinfo{person}{Jian-Yun Nie}.} \bibinfo{year}{2026}\natexlab{}.
\newblock \showarticletitle{{ConvMix}: A Mixed-Criteria Data Augmentation
  Framework for Conversational Dense Retrieval}. In
  \bibinfo{booktitle}{\emph{Proceedings of the AAAI Conference on Artificial
  Intelligence}}, Vol.~\bibinfo{volume}{40}. \bibinfo{pages}{15555--15563}.
\newblock
\urldef\tempurl%
\url{https://doi.org/10.1609/aaai.v40i18.38584}
\showDOI{\tempurl}


\bibitem[Mo et~al\mbox{.}(2024e)]%
        {mo2024leverage}
\bibfield{author}{\bibinfo{person}{Fengran Mo}, \bibinfo{person}{Longxiang
  Zhao}, \bibinfo{person}{Kaiyu Huang}, \bibinfo{person}{Yue Dong},
  \bibinfo{person}{Degen Huang}, {and} \bibinfo{person}{Jian-Yun Nie}.}
  \bibinfo{year}{2024}\natexlab{e}.
\newblock \showarticletitle{How to Leverage Personal Textual Knowledge for
  Personalized Conversational Information Retrieval}. In
  \bibinfo{booktitle}{\emph{Proceedings of the 33rd ACM International
  Conference on Information and Knowledge Management (CIKM)}}.
  \bibinfo{pages}{3954--3958}.
\newblock
\urldef\tempurl%
\url{https://doi.org/10.1145/3627673.3679939}
\showDOI{\tempurl}


\bibitem[Oh et~al\mbox{.}(2024)]%
        {oh2024instructir}
\bibfield{author}{\bibinfo{person}{Hanseok Oh}, \bibinfo{person}{Hyunji Lee},
  \bibinfo{person}{Seonghyeon Ye}, \bibinfo{person}{Haebin Shin},
  \bibinfo{person}{Hansol Jang}, \bibinfo{person}{Changwook Jun}, {and}
  \bibinfo{person}{Minjoon Seo}.} \bibinfo{year}{2024}\natexlab{}.
\newblock \showarticletitle{{INSTRUCTIR}: A Benchmark for Instruction Following
  of Information Retrieval Models}.
\newblock \bibinfo{journal}{\emph{arXiv preprint arXiv:2402.14334}}
  (\bibinfo{year}{2024}).
\newblock


\bibitem[Overwijk et~al\mbox{.}(2022)]%
        {overwijk2022clueweb}
\bibfield{author}{\bibinfo{person}{Arnold Overwijk}, \bibinfo{person}{Chenyan
  Xiong}, {and} \bibinfo{person}{Jamie Callan}.}
  \bibinfo{year}{2022}\natexlab{}.
\newblock \showarticletitle{{ClueWeb22}: 10 Billion Web Documents with Rich
  Information}. In \bibinfo{booktitle}{\emph{Proceedings of the 45th
  International ACM SIGIR Conference on Research and Development in Information
  Retrieval}}.
\newblock


\bibitem[Owoicho et~al\mbox{.}(2023)]%
        {owoicho2023exploiting}
\bibfield{author}{\bibinfo{person}{Paul Owoicho}, \bibinfo{person}{Ivan
  Sekuli{\'c}}, \bibinfo{person}{Mohammad Aliannejadi},
  \bibinfo{person}{Jeffrey Dalton}, {and} \bibinfo{person}{Fabio Crestani}.}
  \bibinfo{year}{2023}\natexlab{}.
\newblock \showarticletitle{Exploiting Simulated User Feedback for
  Conversational Search: Ranking, Rewriting, and Beyond}.
\newblock \bibinfo{journal}{\emph{arXiv preprint arXiv:2304.13874}}
  (\bibinfo{year}{2023}).
\newblock


\bibitem[Park et~al\mbox{.}(2023)]%
        {Park2023}
\bibfield{author}{\bibinfo{person}{Joon~Sung Park}, \bibinfo{person}{Joseph~C.
  O'Brien}, \bibinfo{person}{Carrie~J. Cai}, \bibinfo{person}{Meredith~Ringel
  Morris}, \bibinfo{person}{Percy Liang}, {and} \bibinfo{person}{Michael~S.
  Bernstein}.} \bibinfo{year}{2023}\natexlab{}.
\newblock \showarticletitle{Generative Agents: Interactive Simulacra of Human
  Behavior}. In \bibinfo{booktitle}{\emph{Proceedings of the 36th Annual ACM
  Symposium on User Interface Software and Technology (UIST)}}.
\newblock


\bibitem[Qu et~al\mbox{.}(2020)]%
        {qu2020open}
\bibfield{author}{\bibinfo{person}{Chen Qu}, \bibinfo{person}{Liu Yang},
  \bibinfo{person}{Cen Chen}, \bibinfo{person}{Minghui Qiu},
  \bibinfo{person}{W~Bruce Croft}, {and} \bibinfo{person}{Mohit Iyyer}.}
  \bibinfo{year}{2020}\natexlab{}.
\newblock \showarticletitle{Open-retrieval conversational question answering}.
  In \bibinfo{booktitle}{\emph{Proceedings of the 43rd International ACM SIGIR
  conference on research and development in Information Retrieval}}.
  \bibinfo{pages}{539--548}.
\newblock


\bibitem[{Qwen Team}(2025)]%
        {qwen3emb}
\bibfield{author}{\bibinfo{person}{{Qwen Team}}.}
  \bibinfo{year}{2025}\natexlab{}.
\newblock \showarticletitle{{Qwen3} Embedding: Advancing Text Embedding and
  Reranking Through Foundation Models}.
\newblock \bibinfo{journal}{\emph{arXiv preprint arXiv:2506.05176}}
  (\bibinfo{year}{2025}).
\newblock


\bibitem[Robertson and Zaragoza(2009)]%
        {bm25}
\bibfield{author}{\bibinfo{person}{Stephen Robertson} {and}
  \bibinfo{person}{Hugo Zaragoza}.} \bibinfo{year}{2009}\natexlab{}.
\newblock \showarticletitle{The Probabilistic Relevance Framework: BM25 and
  Beyond}.
\newblock \bibinfo{journal}{\emph{Foundations and Trends in Information
  Retrieval}} \bibinfo{volume}{3}, \bibinfo{number}{4} (\bibinfo{year}{2009}),
  \bibinfo{pages}{333–389}.
\newblock
\showISSN{1554-0669}
\urldef\tempurl%
\url{https://doi.org/10.1561/1500000019}
\showDOI{\tempurl}


\bibitem[Son et~al\mbox{.}(2026)]%
        {son2026rcem}
\bibfield{author}{\bibinfo{person}{Kilho Son}, \bibinfo{person}{Paul Hsu},
  \bibinfo{person}{Cha Zhang}, {and} \bibinfo{person}{Dinei Florencio}.}
  \bibinfo{year}{2026}\natexlab{}.
\newblock \showarticletitle{{RCEM}: Robust Conversational Search EMbedder in
  Distributional Shift}.
\newblock \bibinfo{journal}{\emph{arXiv preprint arXiv:2606.01697}}
  (\bibinfo{year}{2026}).
\newblock


\bibitem[Tsirigotis et~al\mbox{.}(2025)]%
        {tsirigotis2025bixse}
\bibfield{author}{\bibinfo{person}{Christos Tsirigotis},
  \bibinfo{person}{Vaibhav Adlakha}, \bibinfo{person}{Jo{\~a}o Monteiro},
  \bibinfo{person}{Aaron Courville}, {and} \bibinfo{person}{Perouz Taslakian}.}
  \bibinfo{year}{2025}\natexlab{}.
\newblock \showarticletitle{{BiXSE}: Improving Dense Retrieval via
  Probabilistic Graded Relevance Distillation}. In
  \bibinfo{booktitle}{\emph{Conference on Language Modeling (COLM)}}.
\newblock
\newblock
\shownote{arXiv:2508.06781}.


\bibitem[van~den Oord et~al\mbox{.}(2019)]%
        {oord2019representation}
\bibfield{author}{\bibinfo{person}{Aaron van~den Oord}, \bibinfo{person}{Yazhe
  Li}, {and} \bibinfo{person}{Oriol Vinyals}.} \bibinfo{year}{2019}\natexlab{}.
\newblock \bibinfo{title}{Representation Learning with Contrastive Predictive
  Coding}.
\newblock
\newblock
\showeprint[arxiv]{1807.03748}~[cs.LG]


\bibitem[Wang et~al\mbox{.}(2026)]%
        {wang2026skillfeed}
\bibfield{author}{\bibinfo{person}{Tianle Wang}, \bibinfo{person}{Yanghe Zou},
  \bibinfo{person}{Xiang Liu}, \bibinfo{person}{Ziyao Huang},
  \bibinfo{person}{Chenchen Fu}, {and} \bibinfo{person}{Weiwei Wu}.}
  \bibinfo{year}{2026}\natexlab{}.
\newblock \showarticletitle{Beyond Task-Only Matching: Personalized Skill
  Routing with Counterfactual Evaluation}.
\newblock \bibinfo{journal}{\emph{arXiv preprint arXiv:2608.28241}}
  (\bibinfo{year}{2026}).
\newblock


\bibitem[Weller et~al\mbox{.}(2025)]%
        {weller2025promptriever}
\bibfield{author}{\bibinfo{person}{Orion Weller}, \bibinfo{person}{Benjamin
  Van~Durme}, \bibinfo{person}{Dawn Lawrie}, \bibinfo{person}{Ashwin
  Paranjape}, \bibinfo{person}{Yuhao Zhang}, {and} \bibinfo{person}{Jack
  Hessel}.} \bibinfo{year}{2025}\natexlab{}.
\newblock \showarticletitle{Promptriever: Instruction-Trained Retrievers Can Be
  Prompted Like Language Models}. In \bibinfo{booktitle}{\emph{International
  Conference on Learning Representations (ICLR)}}.
\newblock


\bibitem[Yang et~al\mbox{.}(2025)]%
        {yang2025contextual}
\bibfield{author}{\bibinfo{person}{Seunghan Yang}, \bibinfo{person}{Juntae
  Lee}, \bibinfo{person}{Jihwan Bang}, \bibinfo{person}{Kyuhong Shim},
  \bibinfo{person}{Minsoo Kim}, {and} \bibinfo{person}{Simyung Chang}.}
  \bibinfo{year}{2025}\natexlab{}.
\newblock \showarticletitle{Learning Contextual Retrieval for Robust
  Conversational Search}. In \bibinfo{booktitle}{\emph{Proceedings of the 2025
  Conference on Empirical Methods in Natural Language Processing (EMNLP)}}.
  \bibinfo{pages}{12002--12014}.
\newblock
\urldef\tempurl%
\url{https://doi.org/10.18653/v1/2025.emnlp-main.602}
\showDOI{\tempurl}


\bibitem[Yu et~al\mbox{.}(2021)]%
        {yu2021few}
\bibfield{author}{\bibinfo{person}{Shi Yu}, \bibinfo{person}{Zhenghao Liu},
  \bibinfo{person}{Chenyan Xiong}, \bibinfo{person}{Tao Feng}, {and}
  \bibinfo{person}{Zhiyuan Liu}.} \bibinfo{year}{2021}\natexlab{}.
\newblock \showarticletitle{Few-shot conversational dense retrieval}. In
  \bibinfo{booktitle}{\emph{Proceedings of the 44th International ACM SIGIR
  Conference on Research and Development in Information Retrieval}}.
  \bibinfo{pages}{829--838}.
\newblock


\bibitem[Zamani et~al\mbox{.}(2023)]%
        {Zamani2022ConversationalIS}
\bibfield{author}{\bibinfo{person}{Hamed Zamani}, \bibinfo{person}{Johanne~R.
  Trippas}, \bibinfo{person}{Jeffrey Dalton}, {and} \bibinfo{person}{Filip
  Radlinski}.} \bibinfo{year}{2023}\natexlab{}.
\newblock \showarticletitle{Conversational Information Seeking}.
\newblock \bibinfo{journal}{\emph{Found. Trends Inf. Retr.}}
  \bibinfo{volume}{17}, \bibinfo{number}{3--4} (\bibinfo{year}{2023}),
  \bibinfo{pages}{244--456}.
\newblock
\urldef\tempurl%
\url{https://api.semanticscholar.org/CorpusID:246210119}
\showURL{%
\tempurl}


\bibitem[Zeng et~al\mbox{.}(2023)]%
        {zeng2023uia}
\bibfield{author}{\bibinfo{person}{Hansi Zeng}, \bibinfo{person}{Surya
  Kallumadi}, \bibinfo{person}{Zaid Alibadi}, \bibinfo{person}{Rodrigo
  Nogueira}, {and} \bibinfo{person}{Hamed Zamani}.}
  \bibinfo{year}{2023}\natexlab{}.
\newblock \showarticletitle{A Personalized Dense Retrieval Framework for
  Unified Information Access}. In \bibinfo{booktitle}{\emph{Proceedings of the
  46th International ACM SIGIR Conference on Research and Development in
  Information Retrieval}}. \bibinfo{pages}{121--130}.
\newblock
\urldef\tempurl%
\url{https://doi.org/10.1145/3539618.3591626}
\showDOI{\tempurl}


\bibitem[Zhang et~al\mbox{.}(2020)]%
        {zhang2020dpsr}
\bibfield{author}{\bibinfo{person}{Han Zhang}, \bibinfo{person}{Songlin Wang},
  \bibinfo{person}{Kang Zhang}, \bibinfo{person}{Zhiling Tang},
  \bibinfo{person}{Yunjiang Jiang}, \bibinfo{person}{Yun Xiao},
  \bibinfo{person}{Weipeng Yan}, {and} \bibinfo{person}{Wen-Yun Yang}.}
  \bibinfo{year}{2020}\natexlab{}.
\newblock \showarticletitle{Towards Personalized and Semantic Retrieval: An
  End-to-End Solution for E-commerce Search via Embedding Learning}. In
  \bibinfo{booktitle}{\emph{Proceedings of the 43rd International ACM SIGIR
  Conference on Research and Development in Information Retrieval}}.
  \bibinfo{pages}{2407--2416}.
\newblock
\urldef\tempurl%
\url{https://doi.org/10.1145/3397271.3401446}
\showDOI{\tempurl}


\bibitem[Zhang et~al\mbox{.}(2024)]%
        {Zhang2024}
\bibfield{author}{\bibinfo{person}{Zeyu Zhang}, \bibinfo{person}{Xiaohe Bo},
  \bibinfo{person}{Chen Ma}, \bibinfo{person}{Rui Li}, \bibinfo{person}{Xu
  Chen}, \bibinfo{person}{Quanyu Dai}, \bibinfo{person}{Jieming Zhu},
  \bibinfo{person}{Zhenhua Dong}, {and} \bibinfo{person}{Ji-Rong Wen}.}
  \bibinfo{year}{2024}\natexlab{}.
\newblock \showarticletitle{A Survey on the Memory Mechanism of Large Language
  Model based Agents}.
\newblock \bibinfo{journal}{\emph{arXiv preprint arXiv:2404.13501}}
  (\bibinfo{year}{2024}).
\newblock


\end{thebibliography}

\appendix

\begin{figure*}[t]
  \centering
  \footnotesize
  \fbox{\begin{minipage}[t]{0.40\textwidth}
    \textbf{User profile}\par\smallskip
    \begin{tabular}{@{}r@{\hspace{4pt}}p{0.85\linewidth}@{}}
      1. & I practice yoga daily. \\
      2. & I have curly hair that falls just past my shoulders. \\
      3. & I work as a graphic designer for a tech startup. \\
      4. & I enjoy cooking, especially Italian cuisine. \\
      5. & I dream of opening my art gallery someday. \\
      6. & I have two sisters. \\
      7. & I have a small tattoo of a hummingbird on my ankle. \\
      8. & I am 5'7'' tall and have an athletic body. \\
      9. & My eyes are a striking shade of green. \\
      10. & I love painting in watercolor in my free time. \\
      11. & I volunteer at a local animal shelter. \\
      $\triangleright$\,\textbf{12.} & \textbf{I am from San Francisco, California.} \\
      13. & I often travel to Europe for inspiration. \\
      14. & I am an avid reader of historical fiction. \\
      15. & I enjoy hiking on weekends. \\
      16. & I'm 32 years old. \\
      17. & I have a golden retriever dog named Max. \\
      18. & I think I'm funny. \\
      19. & I speak Spanish and French. \\
      20. & I am a passionate advocate for environmental causes. \\
      21. & I have a close-knit group of friends. \\
    \end{tabular}
  \end{minipage}}\hfill
  \fbox{\begin{minipage}[t]{0.555\textwidth}
    \textbf{Conversation}\par\smallskip
    \begin{tabular}{@{}l@{\hspace{5pt}}p{0.82\linewidth}@{}}
      User: & I'm thinking about traveling to Egypt. What is the best time of year to visit there for pleasant weather? \\[2pt]
      System: & The best time to visit Egypt is during winter from October to April, when temperatures are lower. The most popular time for tourists is December and January. Tourism in Egypt is booming this time of the year so make sure you plan in advance. March, April, and September are also good months to travel to Egypt if you plan to avoid the crowd. Most countries need a visa, do you want to know about visas? \\[2pt]
      User: & \textbf{Oh, do I need a visa?} \hfill \textit{current question} \\
    \end{tabular}
    \par\medskip
    \textbf{Human oracle rewrite}\par\smallskip
    Oh, do I need a visa to visit Egypt as an American citizen?
    \par\medskip
    \textbf{Relevant profile statements}\par\smallskip
    Statement~12.
  \end{minipage}}
  \caption{A judged training turn from iKAT 2024, with its user profile, dialogue history, and current question. Only statement~12, marked with $\triangleright$, is annotated as relevant. The human oracle rewrite is not part of the PCDR input and serves only as the distillation teacher.}
  \label{fig:sample}
\end{figure*}

\section{Sample Conversation and User Profile}
\label{app:sample}
Figure~\ref{fig:sample} shows a judged training turn from iKAT 2024, the second turn of the conversation ``Traveling to Egypt''. The user profile contains $21$ statements, of which the annotators mark only statement~$12$ as relevant to the current question. The question ``\textit{Oh, do I need a visa?}'' depends on both sources of context. The dialogue history establishes that the visa is for Egypt, and the profile implies, without stating it, that the user is an American citizen, which determines whether a visa is required. The human oracle rewrite makes both pieces of information explicit. PCDR receives the instruction, the full profile, and the conversation as a single string, as described in Section~\ref{sec:query}, and the oracle rewrite is used only as the distillation teacher of Section~\ref{sec:loss}.

\section{Is PCDR Better than PCQR? An Inconclusive Study}
\label{app:pcdr-vs-pcqr}
Our motivation for PCDR is that dense encoding should be more robust than
Personalized Conversational Query Reformulation (PCQR). We ran a controlled
experiment to test this directly, but the held-out set turned out to be too
small to draw a statistically reliable conclusion. We report it here for
completeness and to motivate future work.

\paragraph{Setup.} We fix the retriever to the raw, un-fine-tuned
base-Qwen3-Embedding index and vary only how the query is built. For PCQR, a
rewriter reads the full conversation and the full user profile and produces a
self-contained personalized query, which is then encoded with the same ad-hoc
instruction template as our other Qwen3 rows; we use the minimal MQ4CS prompt
listed in Appendix~\ref{app:prompts} at temperature $0$. We sweep the rewriter across the
open-weight Qwen3 model sizes of $0.6$B, $1.7$B, $4$B, and $32$B, the frontier model
gpt-4o, and the human oracle rewrite. PCDR instead encodes the conversation and
the full user profile directly, with no generative call. All rows are evaluated
on the same $44$ held-out turns with full-corpus retrieval and NDCG@$3$. To score
all rewrites in a single pass over the corpus, the rewriter rows use exact search
over int8-quantized document embeddings, whereas PCDR is evaluated with the
full-precision index of Section~\ref{sec:experiments}. The quantization changes
scores only slightly: the human oracle rewrite obtains $36.7$ on the int8 index
and $36.3$ on the full-precision index, the latter being the upper bound reported
in Table~\ref{tab:cumulative}.

\paragraph{Significance testing.} For each rewriter we compare its per-turn
NDCG@$3$ against PCDR with a paired Wilcoxon signed-rank test over the $44$ turns,
and apply a Bonferroni correction for the six comparisons, giving a threshold of
$\alpha = 0.05/6 \approx 0.0083$.

\paragraph{Result.} Table~\ref{tab:pcdr-vs-pcqr} reports the comparison; $\Delta$ and $p$ are computed against PCDR, and win/loss/tie counts the turns where the rewriter beats, loses to, or ties PCDR. For reference, the same raw encoder scores $12.9$ on the direct full-profile query and $29.0$ after conversational fine-tuning. Only the same-size $0.6$B
rewriter is significantly worse than PCDR. Every other comparison, including the
$4$B, $32$B, and gpt-4o rewriters and even the human oracle rewrite, is
\emph{not} statistically significant after correction. In other words, on this
data we can neither conclude that PCDR beats PCQR nor the reverse: PCDR is
statistically indistinguishable from strong rewriters and from the human oracle,
while a rewriter of the same size as the PCDR encoder is clearly worse. The point
estimates suggest the reformulate-then-retrieve pipeline needs a rewriter roughly
an order of magnitude larger to match direct encoding, and pays an extra
generative call per turn, but with $44$ queries the test is underpowered and we
do not claim this as a finding.

\paragraph{Future work.} A conclusive comparison needs far more personalized
relevance judgments than the iKAT held-out split provides. A promising direction
is to train PCDR on a larger \emph{synthesized} personalized dataset and then
compare PCDR against PCQR across all iKAT test collections from 2023 to 2025 at once,
where the combined query set would give the statistical power this appendix lacks.

\begin{table}[t]
    \centering
    \caption{PCDR versus PCQR with rewriters of different sizes, in NDCG@$3$ on the $44$ held-out turns. $\Delta$ and $p$ are computed against PCDR with a paired Wilcoxon signed-rank test, and $^{\ast}$ indicates significance after Bonferroni correction. Rewriter rows are scored on an int8-quantized document index.}
    \vspace{-2ex}
    \label{tab:pcdr-vs-pcqr}
    \setlength{\tabcolsep}{4pt}
    \resizebox{\columnwidth}{!}{%
    \begin{tabular}{@{}lcccc@{}}
    \toprule
        Query & NDCG@3 & $\Delta$ & $p$ & w/l/t \\
        \midrule
        \textbf{PCDR} & \textbf{33.1} & --- & --- & --- \\
        \midrule
        Qwen3-$0.6$B rw. & $11.7$ & $-21.4$ & $0.0001^{\ast}$ & $3/23/18$ \\
        Qwen3-$1.7$B rw. & $20.8$ & $-12.3$ & $0.013$ & $8/19/17$ \\
        Qwen3-$4$B rw. & $33.5$ & $+0.4$ & $0.60$ & $16/11/17$ \\
        Qwen3-$32$B rw. & $36.1$ & $+3.0$ & $0.066$ & $21/12/11$ \\
        gpt-4o rw. & $39.5$ & $+6.5$ & $0.10$ & $17/14/13$ \\
        human oracle rw. & $36.7$ & $+3.6$ & $0.50$ & $15/15/14$ \\
        \bottomrule
    \end{tabular}}
\end{table}

\section{Prompts}
\label{app:prompts}
For completeness we list, verbatim, the prompts used by the submitted runs: the two query reformulations (\texttt{rar} and \texttt{MQ4CS\_persq}), the per-turn user-profile extraction prompt, and the response generation prompt. Curly braces denote fields filled at run time; few-shot demonstrations loaded from a data file are abbreviated. These listings reproduce the prompts as they appear in our code, where the user profile is named \texttt{PTKB}.

\subsection{Query reformulation: \texttt{rar}}
The \texttt{rar} reformulation (used by the SPLADE-v3 leg of Run~5) decontextualizes the current question into a self-contained rewrite; only the rewrite is taken as the retrieval query.
\begin{lstlisting}[style=prompt]
For an information-seeking dialog, please help reformulate the question into rewrite that can fully express the user's information needs without the need of context, but also generate an informative response to answer the question. I will give you several example multi-turn dialogs, where each turn contains a question as well as a rewrite and a response that you need to generate.

[few-shot example dialogs omitted]

YOUR TASK (only questions and responses may be given):
Context:
Question: {previous question}
Response: {previous response}
...
Current Question: {current utterance}

Now, you should give me the rewrite and response of the **Current Question** under the **Context**. The output format should always be:
Rewrite: $Rewrite
Response: $Response.
Note that you should always try to rewrite it and generate an informative response. Never ask for clarification or say you don't understand it in the generated rewrite and response. Go ahead!
\end{lstlisting}

\subsection{Query reformulation: \texttt{MQ4CS\_persq}}
The \texttt{MQ4CS\_persq} reformulation (used by the BM25 leg of Run~4) rewrites the current question into a self-contained, personalized query, using the user profile as background knowledge.
\begin{lstlisting}[style=prompt]
# Instruction:
I will give you a conversation between a user and a system. Also, I will give you some background information about the user. You should rewrite the last question of the user into a self-contained query.

# Background Knowledge: 1: {profile statement 1} 2: {profile statement 2} ...
# Context:
user: {previous user utterance}
system: {previous system response}
...
# Please rewrite the following user question: {current utterance}
# Re-written query:
\end{lstlisting}

\subsection{User-profile extraction}
After each turn, this prompt asks gpt-5-mini for any new durable, personally-relevant facts the user revealed, which are then merged into the running user profile.
\begin{lstlisting}[style=prompt]
You maintain a user's Personal Text Knowledge Base (PTKB) for a personalized conversational assistant. A PTKB is a set of concise, first-person, present-tense statements capturing DURABLE, personally-relevant facts about the user that help personalize future answers --- across: preferences / likes & dislikes; goals & plans; occupation & background / life facts; habits & routines; health & physical conditions; skills & abilities; values & concerns; possessions; dietary / lifestyle constraints; and salient ongoing situations (e.g. an approaching deadline).

You are given the user's CURRENT PTKB and the LATEST turn of an ongoing conversation. Output any NEW personal facts the user revealed THIS turn that are (a) about the USER (not the world / the assistant's content), (b) durable or decision-relevant (not one-off chit-chat), and (c) NOT already entailed by the current PTKB.

Rules:
- One atomic fact per line, as "I ..." (first person, present tense, concise).
- Capture DURABLE traits / preferences / conditions --- NOT the user's immediate request or task
  this turn (e.g. "help me find a gift" / "I'm looking for ways to do X" is a TASK, not a fact).
- Do NOT extract the fine-grained CONTENT or WORDING of something being co-designed or built in the
  conversation (the exact phrases of a script/prompt, the individual items in a plan or list, tiny
  settings being tweaked together turn by turn). Those are artifacts of THIS task, not the user.
  Extract a fact ONLY if it would still describe the user OUTSIDE this conversation; prefer ONE
  general durable fact over many hyper-specific ones, and stop once the durable preference is captured.
- Use ONLY what the USER stated or clearly implied --- NEVER the assistant's suggestions,
  inferences, or diagnoses.
- The PTKB below ALREADY includes facts extracted earlier in this conversation. Do NOT restate,
  paraphrase, refine, or re-emit anything already entailed by it.
- If the user corrects/updates a prior fact, output the corrected fact.
- If there is NO new durable personal fact this turn, output exactly: NONE

### Example A (a real durable fact)
Current PTKB: 1. I want to increase my protein intake. 2. I eat dinner late at night.
User: "I'm planning meals, but I have to be careful --- spicy and acidic stuff really sets off my acid reflux."
Assistant: "Sure, here are reflux-friendly high-protein options..."
New PTKB facts:
I have acid reflux.

### Example B (a task / request is NOT a fact)
Current PTKB: 1. I enjoy hiking.
User: "Can you help me find ways to cheer up a friend who's feeling down?"
Assistant: "Here are ideas to support a friend..."
New PTKB facts:
NONE

### Example C (extract the user's fact, NOT the assistant's inference)
Current PTKB: (empty)
User: "My finger really hurts at the base after a hard climb yesterday."
Assistant: "That could be an A2 pulley injury; rest and ..."
New PTKB facts:
I have finger pain at the base after climbing.

### Example D (co-designed wording / micro-details are NOT durable facts)
Current PTKB: 1. I find short phone voice-prompts helpful for snack reminders.
User: "Great --- let's have it say 'Two breaths', then 'Notice the taste', and start with 'Phone away.'"
Assistant: "Done --- here is your 8-second prompt: ..."
New PTKB facts:
NONE

### Now do it.
Current PTKB:
{ptkb}

Latest turn:
User: {utterance}
Assistant: {response}

New PTKB facts (one per line, or NONE):
\end{lstlisting}

\subsection{Response generation}
The retrieval-augmented response is generated from the top-ranked passages with the following prompt.
\begin{lstlisting}[style=prompt]
# Task Description:
You are a helpful, knowledgeable assistant talking with the user in an ongoing conversation. You are given the user's profile (PTKB) and your Search Results for their last question. Write a reply to the user's last question that, above all, SOUNDS NATURAL -- like a real person speaking, not a report, a database export, or a list. Follow ALL of these:
	a. Natural, conversational tone (THE MOST IMPORTANT CRITERION): Write in flowing, warm, human prose. Lead with a direct, helpful answer, then weave the relevant details into sentences. Do NOT produce long structured checklists, multi-section plans, day-by-day schedules, printable templates, or exhaustive bulleted/grocery lists -- EVEN IF the user explicitly asks for one, answer conversationally with the couple of most useful points (a concrete example or two is fine) and keep it brief. It must read like a knowledgeable friend chatting, NOT a document, a template, or a search engine. Answer DIRECTLY, as if you simply know the information -- do NOT keep announcing where it came from. An occasional light "from what I found" is fine, but do NOT overuse it, and NEVER say things like "the search results recommend", "the material I reviewed", "the sources say", "ideas found:", "I found them on Amazon", or "the trail picks I saw", and never use the words "passage", "passages", "reference passage(s)", or "the provided passages/documents/sources". Do not number or cite sources inline, and do not copy text verbatim.
	b. Grounded: Base your answer ONLY on the information in your search results -- no outside knowledge, no invented or assumed facts. If your search didn't cover something the user asked, just say so briefly and naturally rather than guessing.
	c. Answer the actual question: Directly answer what the user LITERALLY asked -- if they ask for specific things (a particular sport, a product name, a number, a recommendation), give concrete specifics rather than only offering to find them. Include the key relevant facts. But do NOT repeat what you already said in earlier turns -- build on it or add something new; if the user re-asks or the conversation circles back, briefly acknowledge what you covered and add new information rather than restating it.
	d. Personalized: Tailor the answer using ONLY the profile facts relevant to this question; ignore the irrelevant ones and never force-fit the profile.
	e. Length: Keep it brief and conversational -- usually ~120-180 words (a short chat reply), and never exceeding ~250 words. A concise, on-point answer is better than a long one.

===TAIL_INSTRUCTION===
Now, please reply to the user's **Last Question** under the **Dialog Context**, considering the **User Profile** and using your **Search Results**. The output format should always be:

Response: $Response

Go ahead!
\end{lstlisting}

\end{document}